\documentclass[11pt]{article}

\usepackage[T1]{fontenc}
\usepackage[utf8]{inputenc}
\usepackage{lmodern}
\usepackage[margin=1in]{geometry}
\usepackage{amsmath,amssymb,amsthm,mathtools}
\usepackage{microtype}
\usepackage[colorlinks=true,linkcolor=blue,citecolor=blue,urlcolor=blue]{hyperref}
\usepackage{enumitem}
\newtheorem{theorem}{Theorem}[section]
\newtheorem{lemma}[theorem]{Lemma}
\newtheorem{proposition}[theorem]{Proposition}

\theoremstyle{definition}
\newtheorem{definition}[theorem]{Definition}
\theoremstyle{remark}
\newtheorem{remark}[theorem]{Remark}
\newcommand{\HH}{\mathcal H}
\newcommand{\LL}{\mathcal L}
\newcommand{\FF}{\mathbb F}

\newcommand{\Tr}{\operatorname{Tr}}
\newcommand{\Min}{\operatorname{Min}}

\newcommand{\eps}{\varepsilon}

\title{Unconditional Lower Bounds
for the Fredman--Khachiyan Algorithms for Monotone Boolean Duality Testing}
\author{Khaled Elbassioni\thanks{\raggedright Khalifa University of Science and Technology, Abu Dhabi, UAE; \texttt{khaled.elbassioni@ku.ac.ae}.}}
\date{}
\begin{document}
\maketitle
\begin{abstract}
We prove unconditional lower bounds on the number of recursive calls made by the two
algorithms of Fredman and Khachiyan (1996) for monotone Boolean duality testing, as a function of the
explicit input size $N$.
For Algorithm A we show a lower bound of $N^{\Omega(\log N)}$ on the dual pairs
$(\HH_k,\LL_k)$ obtained from the read-once family of Gurvich and Khachiyan.
The proof lower-bounds the number of leaves of every binary decision tree computing
the function by the modulus of its multilinear polynomial evaluated at a complex point
of the unit circle. It therefore applies to every choice of branching variables and,
more generally, to the class of two-child variable-restriction procedures whose terminal
calls have polynomially bounded decision-tree complexity.
On this family the number of calls is exactly $N^{\Theta(\log N)}$.
For Algorithm B, a symmetric specialization of the polynomial padding construction of
Kobayashi, Kurita, and Wasa, with full polynomial blocks, yields at least
$q^{(t-1)k-t(t-1)/2}$ calls, where $k=4t^2$ and $q=\Theta(t^2)$; as a function of $N$
this is $N^{\Omega(\sqrt{\log N/\log\log N})}$ along an infinite family.
The same padded family forces Berge multiplication, for every edge ordering, to generate
an intermediate family of size $N^{\Omega(\sqrt{\log N/\log\log N})}$,
where $N$ is now the combined input--output size. %We also give an ordering upper bound showing that Takata's $N^{\Omega(\log\log N)}$ lower bound has the correct exponent scale on his read-once family.
In an appendix we show that applying the padding construction directly to Algorithm A gives
$\Omega(2^n)$ calls on a dual pair obtained from an unsatisfiable $n$-variable formula,
which is only $N^{\Omega(\sqrt{\log N/\log\log N})}$ in terms of the input size.
No complexity-theoretic assumption is used.
The proofs allow every branching-variable choice permitted by the respective algorithm.
The bounds for the original recursive procedures count every invocation separately, without
sharing calls or adding positive stopping tests. 
%The generalized classes and their permitted terminal rules are specified explicitly below and in the appendices.
\end{abstract}

\section{Introduction}
A hypergraph $\HH\subseteq 2^U$ is a finite family of subsets of a finite vertex set $U$.
For a family $\mathcal C$, we denote by $\Min(\mathcal C)$ the family obtained by removing duplicates and all
members properly containing another member. 
We call $\HH$ {\it reduced} if $\HH=\Min(\mathcal H)$.
A transversal of $\HH$ is a subset of $U$ meeting every edge of $\HH$. We denote by
$\Tr(\HH)$ the family of inclusion-minimal transversals of $\HH$.

For $x\in\{0,1\}^U$, and a reduced hypergraph $\HH\subseteq 2^U$ we can associate a unique Boolean function 
\begin{equation}\label{eq:cnfdnf}
 F_{\HH}(x)=\bigwedge_{E\in\HH}\bigvee_{u\in E}x_u=
 \bigvee_{T\in\Tr(\HH)}\bigwedge_{u\in T}x_u.
\end{equation}
Thus a pair $(\HH,\LL)$ with $\HH=\Min(\HH)$ and
$\LL=\Tr(\HH)$ supplies the irredundant positive CNF and DNF of the same
monotone function. Equivalently, the two DNFs with term families $\HH$ and
$\LL$ are dual: 
%in the convention of \cite{FK96}: 
%the DNF with family $\HH$, evaluated at $\bar x$, equals $1-F_{\HH}(x)$.
$F_{\HH}(x)=1-F_{\LL}(\bar x)$.
The {\it monotone Boolean duality testing problem} is to check for a pair of monotone Boolean functions $F_\HH,G_\LL$, given by their irredundant conjunctive normal forms represented by hypergraphs $\HH,\LL\subseteq 2^U$, if they are dual to each other. The {\it monotone Boolean dualization} problem is to compute, for a given monotone Boolean function $F_\HH$, the prime implicates of the dual of $F_\HH$. Equivalently, the latter problem is to compute the minimal transversals of a given hypergraph $\HH$; it reduces to $|\Tr(\HH)|+1$ calls to the search version of duality testing, which returns a missing transversal when the supplied list is incomplete.

Determining the exact complexity of the duality testing problem is an outstanding open question, which has received considerable attention in the literature, e.g.,   \cite{BI95,BM09,E08,EGM02,FK96,KS03,GM14,Lov92,M24,P97}. Polynomial-time duality testing and dualization algorithms are known for some special cases, see, e.g., \cite{BEGK04,BEM10,D97,EG95,GK00,KBEG-IPL07,KBGE07,M24}.
In 1996, Fredman and Khachiyan~\cite{FK96} gave two algorithms for the duality testing problem with quasi-polynomial running time: Algorithm A runs in $N^{O(\log^2 N)}$ time and Algorithm B, which currently stands as the algorithm with best known sequential complexity, runs in $N^{O(\log N/\log\log N)}$ time, where $N$ is the input size. Here we measure $N$ by the incidence size defined in~\eqref{eq:size} below; it is at most quadratic in $|U|+|\HH|+|\LL|+2$ and at least $|\HH|+|\LL|$, so all bounds in this paper are unaffected, up to the constants in the exponents, by the choice between the two measures. A number of practical implementations \cite{KBEG07,H09,P24,SS18,SS23} of both algorithms showed that both algorithms, and in particular the simpler Algorithm A, behave much better than their theoretical upper bounds proved in~\cite{FK96}, and may thus suggest that their running time analyses could be improved. This hypothesis is supported by the fact that, to the best of our knowledge no non-trivial lower bounds were known on the complexity of these two algorithms. 

In this paper, we give the first superpolynomial lower bounds on the complexity of both algorithms, measured by the number of recursive calls of the original recursion-tree procedures (see Remark~\ref{def:model}): an $N^{\Omega(\log N)}$-lower bound for Algorithm A and an $N^{\Omega(\sqrt{\log N/\log\log N})}$-lower bound for Algorithm B. Very recently Kobayashi, Kurita and Wasa~\cite{KKW26} showed that no duality testing algorithm with running time $N^{o(\sqrt{\log N/\log\log N})}$ exists, assuming the {\it Exponential Time Hypothesis} (ETH).  Their reduction maps a 3CNF formula with $n$ variables into two hypergraphs, $\HH$ and $\LL$, of subexponential total size $2^{O(n^{2/3}(\log n)^{1/3})}$ such that $\LL \subseteq \Tr(\HH)$ and the original 3CNF formula is satisfiable if and only if $\LL\neq \Tr(\HH)$. It leverages a specialized polynomial padding gadget to meticulously scale and bound the hypergraph sizes, effectively ruling out output-polynomial time algorithms  under ETH. Our (unconditional) lower bound for Algorithm B is inspired by their reduction and in fact builds on their same padding technique. When applying the same technique on Algorithm A, we only get a lower bound of $N^{\Omega(\sqrt{\log N/\log\log N})}$ (see Appendix~\ref{sec:weakLB}). To obtain the stronger lower bound of $N^{\Omega(\log N)}$, we instead analyze the algorithm (in fact, a more general class of algorithms represented by binary decision trees) on the read-once family given by Gurvich and Khachiyan~\cite{GK97} combined with a lower bounding technique based on multi-linear polynomials over the complex numbers. 

Superpolynomial lower bounds are also known for Berge multiplication
\cite{Takata02,Takata07} and other transversal-generation algorithms
\cite{EHR14}. In Section~\ref{sec:berge}, we show that Takata's
$N^{\Omega(\log\log N)}$ exponent scale cannot be improved on his
read-once family by a lower bound valid for every edge ordering:
we exhibit an ordering with $N^{O(\log\log N)}$ running time.
For the symmetric padded family, however, deleting any one edge
creates sufficiently many new minimal transversals to force a
penultimate intermediate family of size
$N^{\Omega(\sqrt{\log N/\log\log N})}$ in every ordering of
Berge multiplication. Here $N$ is the combined input--output
incidence size.

\section{Preliminaries}
We permit the empty family and the empty edge, with
\[
 \Tr(\varnothing)=\{\varnothing\},\qquad
 \Tr(\{\varnothing\})=\varnothing.
\]
The conventions for empty conjunctions and disjunctions are $1$ and $0$,
respectively.

We use the explicit incidence size
\begin{equation}\label{eq:size}
 N=2+|U|+\sum_{E\in\HH}(1+|E|)
              +\sum_{T\in\LL}(1+|T|).
\end{equation}
A standard binary encoding changes this by at most a logarithmic factor
in $|U|+2$, which does not change any asymptotic lower bound below. All logarithms are to base two, except where another base is indicated.

\begin{remark}\label{def:model}
We count recursive invocations, including terminal invocations, of the
procedures in Sections 3 and 4 of \cite{FK96}.
Each invocation removes redundant terms and uses the stopping rules and
decomposition prescribed there. Different recursive calls are counted
separately, even when their inputs are equal or isomorphic.
Variable choice and the order of secondary calls may be arbitrary subject
to the original rules. There is no memoization, merging of calls, or
additional test that accepts a nonterminal dual pair.
\end{remark}

The lower bounds therefore lower-bound sequential time in this model.
They do not assert lower bounds for all algorithms solving duality.
On a dual input, all necessary-condition tests in \cite{FK96} pass.
The original stopping rule for A is $|\HH||\LL|\le1$;
the original stopping rule for B is
$\min\{|\HH|,|\LL|\}\le2$.
Family order can be interchanged by the algorithms; throughout the proofs
we track which family represents the CNF and which represents the DNF.

\begin{lemma}[\cite{FK96}]\label{lem:restrictions}
Suppose $\HH$ and $\LL$ represent the same monotone function $F$ as in
\eqref{eq:cnfdnf}. For a vertex $v$, representations of its restrictions are
\begin{align*}
 \HH_{v=1}&=\Min\{E\in\HH:v\notin E\},&
 \LL_{v=1}&=\Min\{T\setminus\{v\}:T\in\LL\},\\
 \HH_{v=0}&=\Min\{E\setminus\{v\}:E\in\HH\},&
 \LL_{v=0}&=\Min\{T\in\LL:v\notin T\}.
\end{align*}
Both restricted pairs $(\HH_{v=1},\LL_{v=1})$ and $(\HH_{v=0}, \LL_{v=0})$ are dual.
\end{lemma}
For completeness, a proof is given in Appendix~\ref{fullproof:lem:restrictions}.

\section{An $N^{\Omega(\log N)}$-lower bound for Algorithm A}

\subsection{Variable-restriction algorithms}
We consider a broad class of algorithms which includes Algorithm A as a special case.

\begin{definition}[Variable-restriction algorithms with bounded terminal complexity]
\label{def:restriction-class}
Let $\alpha\colon\mathbb N\to[1,\infty)$ be a fixed function.
A duality-testing procedure belongs to $\mathcal R_\alpha$ if, on every
reduced dual input $(\HH,\LL)$ of incidence size $N$ representing
$F=F_{\HH}$, its execution has the following properties.
\begin{enumerate}[label=(\roman*)]
\item Its recursive calls form a finite rooted binary tree. Each node is
associated with a partial assignment $\rho$ to the original vertex
variables and represents the restriction $F|_\rho$; the root has the
empty assignment.
\item At every nonterminal node it chooses any unassigned variable
$x_v$ and executes both children, representing
$F|_{\rho\cup\{x_v=0\}}$ and $F|_{\rho\cup\{x_v=1\}}$.
Variable choice may depend arbitrarily on the input, the current
representations, and the execution history.
\item At every terminal node, the restricted Boolean function
$F|_\rho$ admits a deterministic variable-query decision tree with at
most $\alpha(N)$ leaves. This is a semantic condition on the residual
function; the procedure need not construct that decision tree.
\item Distinct invocations are counted separately, even when their
inputs are equal or isomorphic. Calls are not merged or memoized.
\end{enumerate}
The procedure may use arbitrary preprocessing and simplification that
preserve these properties, and its behavior on nondual inputs is not
otherwise constrained here. We call $\bigcup_{C\ge0}\mathcal R_{2N^C}$
the class with polynomial terminal decision-tree complexity, where
$\mathcal R_{2N^C}$ denotes the choice $\alpha(N)=2N^C$.
\end{definition}

This definition specifies a restriction on the algorithm, not on the
intrinsic complexity of duality testing. An arbitrary acceptance test
for duality need not satisfy condition (iii): the fact that a residual
pair is dual does not imply that its common Boolean function has a
small decision tree. Likewise, the two-child rule in (ii) is essential;
it is not the multiway secondary decomposition used by FK-B.

\begin{lemma}\label{lem:restriction-completion}
Original Algorithm A belongs to $\mathcal R_2$.
More generally, an execution in $\mathcal R_\alpha$ with $L$ terminal
invocations can be completed to a decision tree for $F_{\HH}$ with at
most $\alpha(N)L$ leaves.
\end{lemma}
\begin{proof}
For Algorithm A, Lemma~\ref{lem:restrictions} identifies the two children
with the two restrictions, keeping track of the CNF and DNF roles if
family order is interchanged. On a dual input all descendants are dual,
so no necessary-condition test rejects and neither child can be avoided
by finding a nonduality witness. A terminal pair satisfies
$|\HH'||\LL'|\le1$. An empty family gives a constant function.
Otherwise $\HH'=\{E\}$ and
$\LL'=\Tr(\{E\})=\{\{u\}:u\in E\}$ has one member, forcing
$|E|=1$. Thus every terminal function is constant or a single variable
and has a decision tree with at most two leaves.

For the general assertion, regard each internal call as a query to its
selected variable, and attach to each terminal call a decision tree
for its residual function. Condition (iii) bounds the number of leaves
of each attached tree by $\alpha(N)$. Their sum is at most $\alpha(N)L$,
and the resulting tree computes $F_{\HH}$ on every Boolean input.
\end{proof}

For example, the class permits unrestricted branching heuristics,
redundancy removal, and early stopping whenever the residual function
depends on at most $b(N)$ variables, with terminal budget $2^{b(N)}$.
It also permits stopping when either residual normal form has at most
$c$ terms, for any fixed $c$. Indeed, a DNF with at most $c$ terms on
$v$ variables has a decision tree with at most $(v+1)^c$ leaves: test
the terms successively, stopping at the first false literal in a term
or accepting when a term is fully true. Testing one conjunction takes
at most $v+1$ leaves, and replacing its continuation leaves by the
remaining tests multiplies this upper bound at most $c$ times.
The CNF case follows by complementation. Since $v\le N$, these stopping
rules have polynomial terminal complexity. Omitting one restriction
branch is allowed only if it is still represented by a terminal node
satisfying the same bound; assignments cannot simply be discarded.

\iffalse
We retain the CNF--DNF convention, the explicit input size in
\eqref{eq:size}, and the original recursion-tree model of
Definition~\ref{def:model}. On a different family from the polynomial
padding construction, the lower bound for Algorithm A can be strengthened
to $N^{\Omega(\log N)}$. This does not replace the padding argument of
Theorem~\ref{thm:A} on its own inputs and does not strengthen
Theorem~\ref{thm:B}.
\fi
The family we use for this lower bound is the classical read-once construction of Gurvich and
Khachiyan \cite{GK97}, reproduced in Remark~1 of \cite{FK96}.
The proof below uses the multilinear polynomial of the function to
lower-bound every decision tree computing it. It does not require a
variable-frequency bound to persist under restrictions.

\subsection{The family in the CNF--DNF convention}

Let $f_1(x_1,x_2)=x_1\vee x_2$. Recursively define
\begin{equation}\label{gka:recursion}
 f_{k+1}=(f_k^{(1)}\wedge f_k^{(2)})
                 \vee(f_k^{(3)}\wedge f_k^{(4)}),
\end{equation}
where the four superscripts denote copies on pairwise disjoint sets of
variables. Let $U_k$ be the variable set and $v_k=|U_k|$. Thus
\begin{equation}\label{gka:variables}
 v_k=2\cdot4^{k-1}.
\end{equation}
For a monotone Boolean function $f$, write
$f^d(x)=1-f(\bar x)$ for its monotone dual. Let $\LL_k$ be the family
of minimal true supports of $f_k$, and let $\HH_k$ be the family of
minimal true supports of $f_k^d$. De Morgan's law gives
\[
 f_k(x)=\bigwedge_{E\in\HH_k}\bigvee_{u\in E}x_u
       =\bigvee_{T\in\LL_k}\bigwedge_{u\in T}x_u.
\]
In particular, $F_{\HH_k}=f_k$ and $\LL_k=\Tr(\HH_k)$, as required
by the convention of the preceding sections. Notice that $\LL_k$,
rather than $\HH_k$, is the term family of $f_k$.
We recall the following properties of this construction.

\begin{lemma}\label{gka:sizes}
For every $k\ge1$,
\[
 a_k:=|\LL_k|=2^{2^k-1},\qquad
 b_k:=|\HH_k|=2^{2^k-2}.
\]
Every member of $\LL_k$ has size $2^{k-1}$, and every member of
$\HH_k$ has size $2^k$. Consequently, if
$m_k=|\HH_k|+|\LL_k|$ is the term count and $N_k$ is the incidence
size in \eqref{eq:size}, then
\begin{align}
 m_k&=3\cdot2^{2^k-2},\label{gka:terms}\\
 N_k&=2+2\cdot4^{k-1}+3\cdot2^{2^k-2}
                         +2^{2^k+k-1}.\label{gka:incidences}
\end{align}
In particular,
\begin{equation}\label{gka:sizeconversion}
 \log N_k=\Theta(2^k),\qquad
 v_k=\Theta((\log N_k)^2).
\end{equation}
\end{lemma}
\begin{proof}
The minimal true supports of a conjunction of nonconstant monotone
functions on disjoint variable sets are exactly the unions of one
minimal true support from each factor. Indeed, a true support must
make every factor true, and minimality is equivalent to minimality in
every factor. For a disjunction on disjoint variable sets, its minimal
true supports are exactly the minimal true supports of the individual
factors: a minimal true support makes at least one factor true and
contains no extra variables. These supports are nonempty, so supports
from different factors cannot contain one another.

Applying these facts to \eqref{gka:recursion} yields
$a_1=2$ and $a_{k+1}=2a_k^2$. Dualizing that recurrence gives
\[
 f_{k+1}^d=
 ((f_k^d)^{(1)}\vee(f_k^d)^{(2)})
 \wedge((f_k^d)^{(3)}\vee(f_k^d)^{(4)}),
\]
so $b_1=1$ and $b_{k+1}=(2b_k)^2$.
Taking logarithms and solving the two recurrences gives the asserted
values of $a_k,b_k$. The same disjoint-support descriptions show that
both term lengths double at each step. Their initial values are one
and two, respectively.

The total number of incidences is therefore
\[
 a_k2^{k-1}+b_k2^k=2^{2^k+k-1}.
\]
Adding the vertex count, the term count, and the constant two from
\eqref{eq:size} gives \eqref{gka:incidences}.
Its last summand is a lower bound for $N_k$, and all its summands have
logarithm $O(2^k)$. This proves \eqref{gka:sizeconversion}.
\end{proof}

\subsection{A polynomial evaluation lower bound for decision trees}

For a Boolean function $F$ on $v$ variables, let $P_F$ be its unique multilinear polynomial over $\mathbb C$ agreeing with $F$ on the Boolean cube $\{0,1\}^v$. 
The existence of $P_F$ follows directly by interpolation using the indicators of Boolean assignments. For any $a \in \{0,1\}^v$, let $\mathbf{1}_a(x) = \prod_{i=1}^v [a_i x_i + (1-a_i)(1-x_i)]$. Then the linear combination $P_F(x) = \sum_{a} F(a)\mathbf{1}_a(x)$ is a multilinear polynomial satisfying the condition.

To establish uniqueness, it suffices to show that any multilinear polynomial $Q$ vanishing on the Boolean cube $\{0,1\}^v$ is identically zero. We proceed by induction on $v$. For $v=1$, the polynomial can be written as $Q(x_1) = A + Bx_1$; setting $x_1=0,1$ immediately yields $A=B=0$. For the inductive step, assume the claim holds for $v-1$ variables. We decompose a polynomial on $v$ variables by isolating $x_v$:
\[
Q(x_1, \ldots, x_v) = A(x_1, \ldots, x_{v-1}) + x_v B(x_1, \ldots, x_{v-1}),
\]
where $A$ and $B$ are multilinear polynomials on $v-1$ variables. Evaluating $Q$ on $\{0,1\}^v$ with $x_v=0$ forces $A(a) = 0$ for all $a \in \{0,1\}^{v-1}$, meaning $A \equiv 0$ by the inductive hypothesis. Subsequently evaluating with $x_v=1$ forces $B(a) = 0$ for all $a \in \{0,1\}^{v-1}$, which similarly implies $B \equiv 0$. Hence, $Q$ is identically zero, confirming uniqueness.

\begin{lemma}[Complex evaluation]\label{gka:complex}
Let $z\in\mathbb C$ satisfy $|z|\le1$ and $|1-z|\le1$.
Every Boolean decision tree computing $F$ has at least
$|P_F(z,\ldots,z)|$ leaves.
\end{lemma}
\begin{proof}
A repeated query on a root-to-leaf path can be removed by retaining
the branch forced by the earlier answer. This does not increase the
number of leaves. We may therefore assume no variable is queried
twice on a path.

For a leaf $\lambda$, let $S^1_\lambda$ and $S^0_\lambda$ be the
disjoint sets of variables assigned one and zero on its path. Its
path indicator is the multilinear polynomial
\[
 Q_\lambda(x)=\prod_{i\in S^1_\lambda}x_i
                  \prod_{j\in S^0_\lambda}(1-x_j).
\]
On any Boolean input $b \in \{0,1\}^v$, the execution trace follows a unique path through the tree. Consequently, exactly one leaf indicator evaluates to one, while all other leaf indicators evaluate to zero.
Thus the sum
of $Q_\lambda$ over leaves labeled one agrees with $F$ on the Boolean
cube. By uniqueness of multilinear representation,
\[
 P_F(x)=\sum_{\lambda:\,\text{label}(\lambda)=1}Q_\lambda(x).
\]
Each summand $T_\lambda(z) = Q_\lambda(z,\ldots,z)$ has modulus at most one at the homogeneous assignment $(z,\ldots,z)$ because $|z| \le 1$ and $|1-z| \le 1$. The triangle inequality bounds $|P_F(z,\ldots,z)|$ by the number of one-labeled leaves, and hence by the total number of leaves:
\[
 |P_F(z, \ldots, z)| \le \sum_{\lambda:\,\text{label}(\lambda)=1} |z|^{|S^1_\lambda|} |1 - z|^{|S^0_\lambda|} \le \text{number of one-labeled leaves},
\]
completing the proof of Lemma~\ref{gka:complex}.

\end{proof}

\begin{lemma}[Exponential growth for the read-once family]
\label{gka:growth}
Set $z=(1+i\sqrt3)/2$, where $i^2=-1$, and put
$u_k=P_{f_k}(z,\ldots,z)$. Then
\begin{equation}\label{gka:evaluationbound}
 |u_k|\ge3^{(v_k+4)/12}
 \qquad(k\ge1).
\end{equation}
\end{lemma}
\begin{proof}
The chosen point satisfies $|z|=|1-z|=1$.
For functions on disjoint variable sets, AND is represented by the
product of their multilinear polynomials and OR by $A+B-AB$.
These expressions remain multilinear because the variable sets are
disjoint. Applying this observation at every gate of the read-once
formula gives
\[
 u_1=2z-z^2=\frac32+\frac{\sqrt3}{2}i,
 \qquad u_{k+1}=2u_k^2-u_k^4.
\]
Let $r_k=|u_k|$, so $r_1=\sqrt3$. If $r_k\ge\sqrt3$, the reverse
triangle inequality gives
\[
 r_{k+1}=r_k^2|2-u_k^2|
 \ge r_k^2(r_k^2-2)
 \ge\frac{r_k^4}{3}.
\]
The last inequality is equivalent to $r_k^2\ge3$. Moreover,
$r_k^4/3\ge3>\sqrt3$, so the hypothesis is preserved inductively.
Writing $s_k=\log_3 r_k$, we have $s_1=1/2$ and
$s_{k+1}\ge4s_k-1$. Equivalently,
$s_{k+1}-1/3\ge4(s_k-1/3)$, and thus
\[
 s_k\ge\frac13+4^{k-1}\left(\frac12-\frac13\right)
       =\frac13+\frac{4^{k-1}}6
       =\frac{v_k+4}{12}.
\]
Exponentiation proves \eqref{gka:evaluationbound}.
\end{proof}

\subsection{Transferring the bound to variable-restriction algorithms}
We now give a stronger lower bound for  algorithms in the class $\mathcal R_\alpha$ which includes Algorithm A of \cite{FK96} as a special case.

\begin{theorem}[Stronger lower bound for variable-restriction algorithms]
\label{gka:theorem}
For every $k\ge1$, every execution of a procedure in $\mathcal R_\alpha$ (and original
Algorithm A) on $(\HH_k,\LL_k)$, with $N=N_k$, has at least
\begin{equation}\label{gka:leafbound}
 \frac1{\alpha(N)}\,3^{(v_k+4)/12}
\end{equation}
terminal invocations. In particular, for original Algorithm A (where $\alpha=2$), and more
generally whenever $\alpha(N)=2^{o((\log N)^2)}$ (for instance $\alpha(N)\le2N^C$), its number of
recursive calls is at least
\[
 2^{\Omega(v_k)}
 =2^{\Omega((\log N_k)^2)}
 =N_k^{\Omega(\log N_k)}.
\]
The same asymptotic bound holds for the standard explicit binary
encoding size. On this family the number of calls is, in fact,
$N_k^{\Theta(\log N_k)}$ for every admissible execution satisfying the preceding condition on $\alpha$.
\end{theorem}
\begin{proof}
Let $L_k$ be the number of terminal invocations. Lemma~\ref{lem:restriction-completion}
produces a decision tree computing $f_k$ with at most $\alpha(N)L_k$ leaves.
Apply Lemmas~\ref{gka:complex} and \ref{gka:growth} at
$z=(1+i\sqrt3)/2$ to obtain
\[
 \alpha(N)L_k\ge |P_{f_k}(z,\ldots,z)|\ge3^{(v_k+4)/12}.
\]
This proves \eqref{gka:leafbound}. The total call count is at least
$L_k$, and Lemma~\ref{gka:sizes} gives the conversion to $N_k$.

For the encoding assertion, replacing vertex incidences by binary
vertex identifiers introduces at most a factor
$O(\log(v_k+2))$. Thus if $\widehat N_k$ is that encoding length,
$\log\widehat N_k=\Theta(\log N_k)$, and the asserted exponent scale
is unchanged.

For the upper bound on calls on this particular family, every
nonterminal invocation fixes a variable absent from all its descendants.
A recursion path therefore has at most $v_k$ branching steps. The
binary execution tree has at most $2^{v_k+1}-1$ nodes. Combining this
with the lower bound and \eqref{gka:sizeconversion} proves the final
assertion. No optimality of this upper bound on general inputs is claimed.
\end{proof}

\begin{remark}[Necessity of Complex Evaluation]
\label{rem:complex-necessity}
It is natural to question whether the complex parameter $z$ in Lemma~\ref{gka:complex} could be replaced by a real number. However, restricting $z \in \mathbb{R}$ collapses the lower bound strategy completely. The domain requirements $|z| \le 1$ and $|1-z| \le 1$ intersect on the real line precisely over the interval $[0,1]$. Within this interval, the base polynomial evaluations yield $|P_{f_1}(z,z)| = 2z - z^2 \le 1$. Consequently, the initial term magnitude satisfies $r_1 = |P_{f_1}(z,z)| \le 1$. Plugging any value $r_1 \le 1$ into the lower-bound recurrence inequality $r_{k+1} \ge r_k^2(r_k^2 - 2)$ makes its right-hand side nonpositive and hence gives no positive growth bound.

By extending the evaluation to the complex plane $\mathbb{C}$, the domain broadens into a geometric lens. Choosing the primitive sixth root of unity $z = \frac{1 + i\sqrt{3}}{2}$ allows us to sit precisely on the boundary curves ($|z| = |1-z| = 1$), while utilizing complex phase rotation to produce $P_{f_1}(z,z) = \frac{3 + i\sqrt{3}}{2}$. This gives a starting magnitude of $r_1 = \sqrt{3} > 1$, which acts as the critical catalyst to trigger the double-exponential growth required for Theorem~\ref{gka:theorem}.
\end{remark}

\iffalse
\begin{remark}[Scope and explicitness]\label{gka:scope}
The two input families can be generated directly from
\eqref{gka:recursion} and its dual using unions and Cartesian products
on disjoint supports. The proof of Lemma~\ref{gka:sizes} shows that
the resulting terms are distinct and minimal. Thus the construction
does not presuppose a transversal oracle and can be written down in
time polynomial in its explicit output size.

No ETH or SETH assumption is used. As in Definition~\ref{def:model},
the conclusion concerns the original recursion tree, with separate
invocations counted separately. It does not cover memoization,
sharing isomorphic subproblems, or additional positive stopping rules.

The conclusion does not transfer to Algorithm B: its secondary
decompositions are not the two-child restriction recursion used in
Lemma~\ref{gka:completion}. In particular, this section leaves the
lower bound of Theorem~\ref{thm:B} unchanged.
\end{remark}
\fi
\section{An $N^{\Omega(\sqrt{\log N/\log\log N})}$-lower bound for Algorithm B}

\subsection{Polynomial padding and its elementary properties}\label{sec:polypad}

Let $U=W_1\dot\cup\cdots\dot\cup W_k$ be a union of nonempty blocks of vertices.
Let $\HH'=\{E'_1,\ldots,E'_M\}$ contain every block $W_i$ (that is, $W_i\in\HH'$ for all $i\in[k]$).
Fix a finite field $\FF_q$, distinct points $\pi_1,\ldots,\pi_M\in\FF_q$, and
an injective labeling, separately within each block, by polynomials $p_u\in \FF_q[x]$
of degree less than $t$. In particular, $M\le q$ and $|W_i|\le q^t$.
Put $h_j(u)=p_u(\pi_j)$. For $a=(a_1,\ldots,a_k)\in\FF_q^k$, set
\begin{equation}\label{eq:padding}
 E_{j,a}=E'_j\cup
       \bigcup_{i=1}^k\{u\in W_i:h_j(u)\ne a_i\}.
\end{equation}
Write $\HH_{\rm raw} = \left\{ E_{j,a} : j \in [M], \, a \in \FF_q^k \right\}$ for the family of these edges and
$\HH=\Min(\HH_{\rm raw})$. This is the padding template of \cite{KKW26}.
We include its needed properties with proofs.

\begin{lemma}[One padding family \cite{KKW26}]\label{lem:onefamily}
A set $X\subseteq U$ meets every $E_{j,a}$ for a fixed $j$ if and only if
either $X\cap E'_j\ne\varnothing$, or some block contains $u,v\in X$
with $h_j(u)\ne h_j(v)$. In either case at most two vertices of $X$
suffice to meet every edge in that family.
\end{lemma}

\begin{proof}
This is the hitting characterization underlying the padding
construction of \cite{KKW26}. A vertex in $E'_j$ meets every
$E_{j,a}$, as do two vertices in one block with distinct evaluations.
Conversely, if neither condition holds, choose each $a_i$ to equal
the common evaluation of the selected vertices in $W_i$, choosing
it arbitrarily when none is selected. The resulting padded edge is
missed. These same observations give the size-two witness.
For completeness, the full proof is included in
Appendix~\ref{fullproof:lem:onefamily}.
\end{proof}

A \emph{full selector} chooses exactly one vertex from each block.

\begin{lemma}[Selectors and small transversals \cite{KKW26}]\label{lem:small}
A set with at most one vertex per block is a transversal of $\HH$
if and only if it is a full selector meeting every edge of $\HH'$.
Each such selector is a minimal transversal. Every minimal transversal which contains
two vertices in one block has size at most $2t$.
\end{lemma}

\begin{proof}
The selector characterization follows from
Lemma~\ref{lem:onefamily}, because $\HH'$ contains every block.
Removing a selected vertex makes the resulting set miss a base
block edge and therefore some edge of its padded family.

The bound for nonselector minimal transversals is the polynomial
collision argument of \cite{KKW26}. Two distinct labels agree at
at most $t-1$ evaluation points. Their two vertices meet every
other padding family, while each exceptional family requires at
most two additional vertices from the given transversal.
Minimality gives $|X|\le2+2(t-1)=2t$.
For completeness, the full proof is included in
Appendix~\ref{fullproof:lem:small}.
\end{proof}

\subsection{A symmetric padding family}

%The argument for Algorithm B uses a different specialization of the same template.
%There is no claim here that arbitrary SAT instances or truncated polynomial
%label sets have the symmetry required by the following proof.

Fix $t\ge2$, set $k=4t^2$, and take $q$ to be the smallest power of two that is at
least $k+1$. Thus
\[
 k+1\le q<2(k+1),\qquad q=\Theta(t^2).
\]
Choose a field of order $q$. For intermediate parameters $m,r$ and any
distinct evaluation points $\pi_0,\pi_1,\ldots,\pi_m$, put
\[
 W_i=\{(i,p):p\in\FF_q[x],\ \deg p<r\},\qquad |W_i|=q^r.
\]
Use the original edges
\[
 E'_0=\varnothing,\qquad E'_i=W_i\quad(1\le i\le m)
\]
and the padding rule \eqref{eq:padding}, with indices $j=0,\ldots,m$:
\begin{equation}\label{eq:padding2}
 E_{j,a}=E'_j\cup
       \bigcup_{i=1}^m\{(i,p)\in W_i:p(\pi_j)\ne a_i\},
\qquad a\in\FF_q^m.
\end{equation}
Let $\HH=\HH_{m,r}$ be the reduced padded family on $U=W_1\dot\cup\cdots\dot\cup W_m$ and
$\LL=\LL_{m,r}=\Tr(\HH_{m,r})$.
Dependence on $q$ and on the evaluation points is suppressed in notation.

\begin{lemma}[Sizes and elementary transversals]\label{lem:Bsmall}
There is no transversal with at most one vertex in each block.
Every member of $\LL_{m,r}$ has size at most $2r$, and
\[
 |\LL_{m,r}|\le(mq^r+1)^{2r+1}.
\]
For each block, every pair of polynomials with nonzero constant difference
is a minimal transversal. In particular,
\[
 |\LL_{m,r}|\ge m q^{r-1}\binom q2.
\]
\end{lemma}
\begin{proof}
Given a set $X$ with at most one vertex per block, choose the $a_i$ in the
$j=0$ family to equal the selected evaluations, (that is, if $(i,p)\in X$, choose $a_i=p(\pi_0)$); the empty original edge
contributes no vertex, so this padded edge is missed: $E_{0,a}\cap X=\varnothing$.
The size bound is the collision argument in Lemma~\ref{lem:small},
with $t$ replaced by $r$. Enumerating all subsets of size at most $2r$
gives the stated upper bound.

If two polynomials differ by a nonzero constant, their evaluations differ
at every point. Their two vertices in the same block therefore meet
every padded family by Lemma~\ref{lem:onefamily}. A singleton is not a
transversal, so the pair is minimal. Fixing the $r-1$ nonconstant
coefficients and choosing two distinct constant coefficients gives
$q^{r-1}\binom q2$ different pairs per block.
\end{proof}

\begin{lemma}[Irredundancy and edge frequencies]\label{lem:Bcounts}
If $m\ge3$ and $r\ge2$, every distinct raw padded edge is minimal.
Furthermore,
\[
 |\HH_{m,r}|=q^m+m q^{m-1},
\]
and the frequency of any vertex on this side (that is, as in~\cite{FK96}, the fraction of members of the family that contain it) is
\begin{equation}\label{eq:Hfreq}
 \alpha_{\HH}(v)=
 1-\frac{q+m-1}{q(q+m)}>1-\frac1q.
\end{equation}
\end{lemma}
\begin{proof}
For a block $W_i$, write
$C_i(j,a)$ for the evaluation fiber $\{(i,p):p(\pi_j)=a\}$. It has size $q^{r-1}$.
For $j\ne h$, the evaluation map \(
\Phi \colon \FF_q[x] \to \FF_q^2\) defined by \( \Phi(p) = \bigl(p(\pi_j), \, p(\pi_h)\bigr)\)
is surjective (onto $\FF_q^2$):
an affine polynomial interpolates any two prescribed values, and
$r\ge2$. Hence $C_i(j,a)$ is not contained in $C_i(h,b)$. 
\iffalse
To prove this from first principles, let $(a, b) \in \FF_q^2$ be an arbitrary pair of target field values. Because $r \ge 2$, the vertex block $W_i$ contains all polynomials over $\FF_q$ of degree strictly less than $r$, which explicitly includes the class of affine polynomials of the form $p(x) = c_1 x + c_0$. Setting up the interpolation conditions yields the linear system:
\[
\begin{cases} 
c_1 \pi_j + c_0 = a \\ 
c_1 \pi_h + c_0 = b 
\end{cases}
\]
Subtracting the second equation from the first isolates the linear coefficient:
\[
c_1 (\pi_j - \pi_h) = a - b.
\]
Since the selected evaluation points are distinct ($\pi_j \neq \pi_h$), the difference term $(\pi_j - \pi_h)$ is a non-zero element of $\FF_q$ and therefore possesses a unique multiplicative inverse. We solve for the coefficients uniquely:
\[
c_1 = \frac{a - b}{\pi_j - \pi_h}, \qquad c_0 = a - \left(\frac{a - b}{\pi_j - \pi_h}\right)\pi_j.
\]
The resulting affine polynomial $p(x) = c_1 x + c_0$ belongs to $W_i$ and satisfies $p(\pi_j) = a$ and $p(\pi_h) = b$ simultaneously. Because such a polynomial can be constructed for any choice of $(a, b)$, the mapping $\Phi$ is surjective.
\fi
The complement of a padded edge in block $i$ is either $C_i(j,a_i)$,
or the empty set when the original edge is $W_i$:
\begin{equation}\label{eq:e1}
W_i \setminus E_{j,a} = (W_i \setminus E'_j) \cap \{(i,p): p(\pi_j) = a_i\}.
\end{equation}
We now show that $\HH_{m,r} = \Min(\HH_{\rm raw})$. Suppose for contradiction that an edge from family $j$ is contained within an edge from a different family $h$, such that $E_{j,a} \subseteq E_{h,b}$ for distinct indices $j \neq h$. Then
\(
W_i \setminus E_{h,b} \subseteq W_i \setminus E_{j,a}.
\)
Choose an index $i\not\in\{j,h\}$, possible since $m\ge3$. Then by~\eqref{eq:e1},
$W_i \setminus E_{h,b} = C_i(h, b_i)$ and $W_i \setminus E_{j,a} = C_i(j, a_i).$
Substituting these fibers back into the restricted containment expression forces the contradiction
\(
C_i(h, b_i) \subseteq C_i(j, a_i).
\)
Next, consider two raw padded edges belonging to the same family index $j$, and suppose $E_{j,a} \subseteq E_{j,b}$. Taking the complement and restricting the inclusion to any block with index $i \neq j$ yields:
\(
C_i(j, b_i) \subseteq C_i(j, a_i).
\)
But $|C_i(j, b_i)| = |C_i(j, a_i)| = q^{r-1}$ then forces
$C_i(j, b_i) = C_i(j, a_i)$ which in turn implies $b_i = a_i$.
Since this equality holds for every $i \neq j$, the target vector components match exactly ($a_i = b_i$ for all $i \neq j$). When $j\ge1$, the remaining coordinate $a_j$ is immaterial because the original edge is $W_j$. When $j=0$, all block coordinates have already been compared.
Hence $E_{j,a}=E_{j,b}$.

The empty-edge family has $q^m$ distinct edges. Each block-edge family
has $q^{m-1}$, since its own block coordinate is immaterial.
Different families have no common edge, by the preceding argument. This establishes the claimed size of $\HH_{m,r}$.
For $v=(i,p)\in W_i$, exactly $q^{m-1}$ edges in the empty-edge family omit $v$ (since we must have $a_i=p(\pi_0)$).
The block-$i$ family omits none, and each of the other $m-1$ block families
has $q^{m-2}$ edges omitting $v$ (since, for $j\neq i$, \(a_{j}\) is immaterial and $a_i=p(\pi_j)$). Dividing their sum by the total count gives \eqref{eq:Hfreq}.
\end{proof}

\begin{lemma}[Intersection preservation under vertex shift]\label{lem:shift}
Let $\tau_g \colon U \to U$ be the induced vertex permutation defined by $\tau_g(i, p) = (i, p + g)$, where $g$ is a fixed polynomial of degree less than $r$. For any subsets of vertices $T, E \subseteq U$, the image of their intersection satisfies
\[
\tau_g(T) \cap \tau_g(E) = \tau_g(T \cap E).
\]
In particular, $\tau_g(\HH_{m,r})$ and $\tau_g(\LL_{m,r})$ are permutations of $\HH_{m,r}$ and $\LL_{m,r}$, respectively.
\end{lemma}
\begin{proof}
The map $\tau_g$ is a bijection with inverse $\tau_{-g}$ and hence
preserves intersections. It preserves each base block and maps
$E_{j,a}$ to $E_{j,a'}$, where $a'_i=a_i+g(\pi_j)$.
It therefore permutes the padded hypergraph. A hypergraph
automorphism preserves hitting and inclusion-minimality, so it
also permutes $\LL_{m,r}=\Tr(\HH_{m,r})$. For completeness, the full proof is included in
Appendix~\ref{fullproof:lem:shift}.
\end{proof}

\begin{lemma}[Transversal frequencies]\label{lem:Lfreq}
Every vertex $v$ of $\LL_{m,r}$ has positive frequency  $\alpha_{\LL}(v)$ satisfying
\[
 \alpha_{\LL}(v)\le\frac{2r}{q^r}.
\]
For $r\ge2$ this is at most $4/q^2$.
\end{lemma}
\begin{proof}
By Lemma~\ref{lem:shift}, addition of a fixed polynomial of degree less than $r$
permutes the vertices in each block, and thus permutes the entire padded hypergraph (where permuted edges stay within their own block family), and hence its minimal
transversals. These translations act transitively on a block.
Every vertex in that block therefore has the same degree in $\LL$.
The sum of these degrees is
$\sum_{T\in\LL}|T\cap W_i|\le2r|\LL|$, by Lemma~\ref{lem:Bsmall}.
Division by $q^r|\LL|$ proves the upper bound.
The constant-difference pairs in that lemma show positivity.
Finally $2r/q^r$ is nonincreasing for integers $r\ge2$ when $q\ge2$,
since the ratio of consecutive values is $(r+1)/(rq)\le1$.
\end{proof}

\subsection{The exact secondary restriction in FK-B}

For a dual pair write
$V=|\HH||\LL|$, and let
\[
 \eps(V)=1/\chi(V),\qquad \chi(V)^{\chi(V)}=V.
\]
Consider a selected vertex with
$\alpha_{\HH}(v)>\eps(V)$ and
$0<\alpha_{\LL}(v)\le\eps(V)$.
Rule (ii) of Section 4 of \cite{FK96} first solves the $x_v=0$ restriction,
and then makes secondary calls indexed by the edges $E\in\HH$
containing $v$. If the family order is reversed, rule (i) has the same
meaning. The following identifies these calls in the CNF--DNF convention.

\begin{lemma}[Secondary child \cite{FK96}]\label{lem:secondary}
Let $J=E\setminus\{v\}$ for an edge $E$ containing $v$.
The secondary pair, up to ordering its two sides, is
\begin{align}
 \HH^E_v&=\Min\{A\setminus J:A\in\HH,\ v\notin A\},\label{eq:secH}\\
 \LL^E_v&=\Min\{T\setminus\{v\}:T\in\LL,\ v\in T,\ T\cap J=\varnothing\}.
 \label{eq:secL}
\end{align}
It represents the restriction $x_v=1$ and $x_u=0$ for all $u\in J$.
\end{lemma}
For completeness, a proof is given in Appendix~\ref{fullproof:lem:secondary}.

\begin{lemma}[Self-reproduction]\label{lem:self}
At the pair $(\HH_{m,r},\LL_{m,r})$ with $m\ge3,r\ge2$, let $v=(i,p_v)$ lie
in block $i$. For each of the $q^{m-1}$ edges from the block-$i$ padding
family, the secondary child of Lemma~\ref{lem:secondary} is isomorphic
to $(\HH_{m-1,r-1},\LL_{m-1,r-1})$ on the remaining evaluation points.
\end{lemma}
\begin{proof}
The edge has the form
\[
 E_{i,a}=W_i\cup
       \bigcup_{\ell\ne i}\{(\ell,p):p(\pi_i)\ne a_\ell\}.
\]
Its secondary restriction fixes $x_v$ to one and $x_u$ to zero for all
$u\in W_i\setminus\{v\}$. In each remaining block it leaves exactly
\[
 D_\ell=\{(\ell,p):p(\pi_i)=a_\ell\}.
\]
Polynomial division by $x-\pi_i$ gives a bijection
\[
 \{g:\deg g<r-1\}\longrightarrow D_\ell,\qquad
 g\longmapsto a_\ell+(x-\pi_i)g.
\]
All edges of the block-$i$ family are satisfied by setting $x_v=1$ and disappear.
For a different index $j$, its original edge is empty or a different
block. Hence an edge in this family $E_{j,a}$ contains $v$ precisely when $a_i\ne p_v(\pi_j)$.
The edges surviving the restriction are exactly those with $a_i=p_v(\pi_j)$.
For such a surviving edge $E_{j,b}$ (where $j \neq i$), its contribution in a remaining block $\ell \neq i$ is given by the intersection $E_{j,b} \cap D_\ell$. If $j = \ell$, the padding rule dictates that $E_{\ell,b}$ contains the entire block $W_\ell$, so that $E_{\ell,b} \cap D_\ell = D_\ell$. 
Otherwise, if $j \neq \ell$, the contribution is restricted to the padding condition, yielding exactly the vertices $p \in D_\ell$ for which $p(\pi_j) \neq b_\ell$.
Under the displayed bijection, the evaluation of $p$ at a remaining point $\pi_j$ (where $j \neq i$) expands to:
\[
 p(\pi_j) = a_\ell + (\pi_j - \pi_i)g(\pi_j).
\]
Because the evaluation points are distinct ($\pi_j \neq \pi_i$), the multiplier $(\pi_j - \pi_i)$ is guaranteed to be a nonzero element of $\FF_q$, making it uniquely invertible. 

The original padding inequality $p(\pi_j) \neq b_\ell$ therefore transforms directly into:
\[
 a_\ell + (\pi_j - \pi_i)g(\pi_j) \neq b_\ell \iff g(\pi_j) \neq \frac{b_\ell - a_\ell}{\pi_j - \pi_i}.
\]
By setting the new threshold for the lower-degree space to be $b'_\ell = (b_\ell - a_\ell)(\pi_j - \pi_i)^{-1}$, the condition becomes exactly $g(\pi_j) \neq b'_\ell$. Since $a_\ell$, $\pi_i$, and $\pi_j$ are fixed parameters, the mapping $b_\ell \mapsto b'_\ell$ constitutes an affine transformation with a nonzero slope over $\FF_q$. Consequently, as the old threshold ranges over $\FF_q$, the corresponding threshold for $g(\pi_j)$ ranges over all of $\FF_q$, preserving the exact edge infrastructure of the reduced family.
Thus all surviving CNF edges, including the empty-base-edge family
at $\pi_0$, are precisely the raw construction with block $i$ and
point $\pi_i$ removed and degree bound $r-1$.
Reducing both families gives its canonical reduced CNF.
By Lemma~\ref{lem:secondary}, the child DNF is its minimal transversal
family, proving the isomorphism on both sides.
\end{proof}

\subsection{Forcing the decomposition at every tracked node}

\begin{lemma}\label{lem:threshold}
For all sufficiently large $t$, at every state
\[
 (m,r)=(k-h,t-h),\qquad 0\le h\le t-2,
\]
every vertex satisfies
\[
 0<\alpha_{\LL}(v)\le\eps(V)<\alpha_{\HH}(v).
\]
(Here, $\HH=\HH_{k-h,t-h}$ and $\LL=\LL_{k-h,t-h}$.) Neither family has at most two members, and every necessary-condition
test passes. Therefore original FK-B executes the secondary decomposition
of Lemma~\ref{lem:secondary}, for every permitted variable choice.
\end{lemma}
\begin{proof}
All these states have $m\ge k-t+2\ge3$ and $r\ge2$.
They are dual by definition. Lemmas~\ref{lem:Bsmall} and
\ref{lem:Bcounts} give more than two members in each family, so the
small-family stopping rule does not apply. They also show, uniformly,
\[
 \log V
 \le \log(q^m+m q^{m-1})+
          (2r+1)\log(mq^r+1)
 =O(t^2\log t),
\]
using $m\le4t^2$, $r\le t$, and $q=\Theta(t^2)$.
For $V\ge4$, $\chi(V)\ge2$ and
$\log V=\chi(V)\log\chi(V)\ge\chi(V)$. Hence
\[
 \eps(V)\ge1/\log V=\Omega(1/(t^2\log t)),
 \qquad \eps(V)\le1/2.
\]
Lemma~\ref{lem:Lfreq} gives
$0<\alpha_{\LL}(v)\le4/q^2=O(t^{-4})$,
which is smaller than this lower bound on $\eps(V)$ for all sufficiently
large $t$. Lemma~\ref{lem:Bcounts} gives
$\alpha_{\HH}(v)>1-1/q>1/2$.
These statements hold uniformly over all remaining evaluation-point
sets. They also hold when the two family names have been interchanged.
The first test of FK-B's step 5, or its symmetric second test, therefore
selects the stated decomposition.
\end{proof}

\begin{theorem}[Lower bound for Algorithm B]\label{thm:B}
For all sufficiently large $t$, every execution of original FK-B on
$(\HH_{k,t},\LL_{k,t})$, where $k=4t^2$ and $q$ is chosen above,
makes at least
\[
 q^{(t-1)k-t(t-1)/2}
\]
recursive calls. For the explicit input size $N$, this is
\[
 N^{\Omega(\sqrt{\log N/\log\log N})}
\]
along this infinite family.
\end{theorem}
\begin{proof}
At a tracked state $(m,r)$ with $r\ge2$, Lemma~\ref{lem:threshold}
forces the required rule for every possible selected vertex $v$.
All subproblems of a dual pair are dual in the decomposition in question,
so the first branch cannot return a nonduality witness that would avoid
the secondary calls. By Lemma~\ref{lem:Bcounts}, all $q^{m-1}$
block-edge terms associated with the block of $v$ survive redundancy
removal and are distinct. Each corresponding secondary call is a copy
of the next state $(m-1,r-1)$ by Lemma~\ref{lem:self}.

Retain only these children in the execution tree. There are at least
$q^{k-h-1}$ retained children per retained node at level $h$,
for $0\le h\le t-2$. The uniformity in Lemma~\ref{lem:threshold}
applies regardless of which block was selected or which evaluation
points remain. The number of invocations at level $t-1$ is therefore
at least
\[
 \prod_{h=0}^{t-2}q^{k-h-1}
 =q^{(t-1)(k-1)-(t-1)(t-2)/2}
 =q^{(t-1)k-t(t-1)/2}.
\]
We need no assertion about how those degree-one calls subsequently behave.
The model counts them separately, even when they are isomorphic.
Ignoring all first branches and all other secondary calls only decreases
the count.

To bound the input size, write $v=kq^t$.
Its reduced hypergraph has exactly $q^k+kq^{k-1}$ edges.
Its transversal family has at most $(v+1)^{2t+1}$ members, each of
size at most $2t$. Therefore
\[
 N\le 2+v+(v+1)(q^k+kq^{k-1})
                   +(2t+1)(v+1)^{2t+1}.
\]
Also $N\ge q^k$. With $k=4t^2$ and $q=\Theta(t^2)$ these inequalities give
\[
 \log N=\Theta(t^2\log t).
\]
The logarithm of the call-count lower bound is
$\Omega(t^3\log t)$. Since
\[
 \log\log N=\Theta(\log t),\qquad
 \sqrt{\frac{\log N}{\log\log N}}=\Theta(t),
\]
this proves the stated bound. Binary vertex labels introduce only a
factor $O(\log(v+2))$ in the encoding size and leave the conclusion unchanged.
\end{proof}

\begin{remark}[Why the decision-tree argument does not transfer to B]
\label{rem:B-decision-tree}
Lemma~\ref{lem:restriction-completion} uses an exhaustive partition of
Boolean inputs: every internal call branches into both restrictions
of one variable. Algorithm B instead uses the duality established by
its first recursive call to simplify the remaining verification
\cite[Section~4, equations (4.1)--(4.5)]{FK96}. Its secondary
restrictions need not cover the complementary half of the Boolean
cube, even together with the main restriction.

For an explicit example, consider the eight-variable member
\[
 f_2=(x_1\vee x_2)(x_3\vee x_4)
           \vee(x_5\vee x_6)(x_7\vee x_8),
 \qquad
 f_2^d=(x_1x_2\vee x_3x_4)(x_5x_6\vee x_7x_8).
\]
Run B on the ordered pair of DNFs $(f_2,f_2^d)$, that is, the term
families $(\LL_2,\HH_2)$, and select $x_1$. Its frequency in $f_2$
is $1/4$, and the volume is $8\cdot4=32$. Since $4^4>32$,
$\chi(32)<4$, so $1/4<\eps(32)$ and rule (i) applies.
The main call represents $f_2|_{x_1=0}$. The two terms of $f_2^d$
containing $x_1$, with $x_1$ removed, are
$J_1=\{x_2,x_5,x_6\}$ and $J_2=\{x_2,x_7,x_8\}$.
The two secondary calls therefore represent, up to taking the dual
and interchanging the family order, the restrictions
\[
 \begin{array}{ll}
 x_1=1,&x_2=x_5=x_6=0,\\
 x_1=1,&x_2=x_7=x_8=0.
 \end{array}
\]
Both restricted functions simplify to $x_3\vee x_4$.
None of these three subcubes contains an assignment with
$x_1=x_2=1$. On that uncovered subcube,
\[
 f_2=(x_3\vee x_4)\vee(x_5\vee x_6)(x_7\vee x_8),
\]
which is nonconstant. Thus the uncovered inputs cannot simply be
represented by a constant decision-tree leaf. B verifies duality on
them through the identity proved in its main call, rather than by
recursively evaluating the common Boolean function there.

Consequently, completing B's terminal calls to small decision trees
does not give a decision tree computing the original function with
only a polynomial increase in leaves. Such a conversion would need
an additional argument accounting for the uncovered regions and
the reuse of the main call's result. Indeed, a polynomial-factor
conversion on this family would imply
$N^{\Omega(\log N)}$ calls by Lemmas~\ref{gka:complex} and
\ref{gka:growth}, contradicting B's general upper bound
$N^{O(\log N/\log\log N)}$. This obstruction to the conversion does
not exclude a smaller improvement of the lower bound for B by a
different argument.
\end{remark}

\section{Berge multiplication: limits of the read-once example and a stronger padding bound}
\label{sec:berge}

The same symmetric padding family also gives a lower bound for
Berge multiplication, or the sequential method. Here the relevant
measure is the size of an intermediate transversal family, rather
than a recursion tree. We first show why the read-once family used
by Takata does not yield an asymptotically stronger lower bound
that holds for every edge ordering, and then obtain such an
improvement from the padded family.

\subsection{The sequential method and its intermediate families}

For an ordering $\sigma=(e_1,\ldots,e_M)$ of a reduced hypergraph
$\HH$, write $\HH_j=\{e_1,\ldots,e_j\}$, including
$\HH_0=\varnothing$. Ordinary Berge multiplication explicitly
maintains the families
\begin{equation}\label{berge:recurrence}
 \mathcal T_0=\{\varnothing\},\qquad
 \mathcal T_j=\Min\{T\cup\{u\}:T\in\mathcal T_{j-1},\ u\in e_j\}
             =\Tr(\HH_j).
\end{equation}
Define its intermediate width under this ordering by
\[
 W(\HH,\sigma)=\max_{0\le j\le M}|\Tr(\HH_j)|.
\]
Explicitly generating the intermediate families takes at least
$W(\HH,\sigma)$ time. Representations storing the whole family
also require at least this many set records at some stage.
Conversely, on a vertex set of size $v$, an update produces at most
$vW(\HH,\sigma)$ candidates. Pairwise inclusion tests give a
straightforward implementation using
$O(Mv^3W(\HH,\sigma)^2)$ time, up to harmless representation
overheads. Thus a width upper bound gives a time upper bound
polynomial in the width and the explicit input size.

Throughout this section $N$ is the combined input--output incidence
size of $(\HH,\Tr(\HH))$, as in \eqref{eq:size}. The lower bound
below allows every permutation of the edges, including an ordering
selected adaptively from the preceding computation. It concerns
the explicit sequential method \eqref{berge:recurrence}, not
methods that avoid generating full prefix transversal families.

\subsection{An ordering upper bound for Takata's read-once family}

For families on disjoint vertex sets define
$\mathcal A\star\mathcal B=\{a\cup b:a\in\mathcal A,\ b\in\mathcal B\}$.
Use the standard form of Takata's family \cite{Takata02,Takata07}:
\begin{equation}\label{berge:takata-family}
 G_0=\{\{u\}\},\qquad
 G_h=(A\cup B)\star(C\cup D),
\end{equation}
where $A,B,C,D$ are vertex-disjoint copies of $G_{h-1}$.
It has the same alternating disjoint-support structure as the
Gurvich--Khachiyan family; its base hypergraph is a singleton
edge rather than the two-vertex edge used for $\HH_1$ above.
Direct induction gives
\begin{equation}\label{berge:takata-counts}
 |V(G_h)|=4^h,\quad |G_h|=2^{2(2^h-1)},\quad
 b_h:=|\Tr(G_h)|=2^{2^h-1},
\end{equation}
and every edge and minimal transversal has size $2^h$.
Indeed, the edge counts satisfy $g_h=4g_{h-1}^2$, while
$b_h=2b_{h-1}^2$, with $g_0=b_0=1$; both set sizes double.
Takata proved that every edge ordering yields an intermediate
family of size $N_h^{\Omega(\log\log N_h)}$, where $N_h$ is the
combined incidence size for $G_h$ and its transversal family.

\begin{proposition}[The exponent scale is tight on Takata's family]
\label{berge:takata-upper}
There is an explicit recursive ordering $\sigma_h$ for which
\[
 W(G_h,\sigma_h)\le2^{h2^h}
                  =N_h^{O(\log\log N_h)}.
\]
Consequently, ordinary Berge multiplication has an implementation
running in $N_h^{O(\log\log N_h)}$ time on this ordering. Together
with Takata's lower bound, the best-order intermediate width on
this family is $N_h^{\Theta(\log\log N_h)}$.
\end{proposition}
\begin{proof}
Order $G_0$ in its unique way. At level $h$, concatenate the
recursive orderings of $A$ and $B$ to order $X=A\cup B$, and
similarly order $Y=C\cup D$. Enumerate $X\star Y$ row by row:
for each edge $e$ of $X$, process $e\cup f$ for all edges $f$ of
$Y$ in their order. All these unions are distinct, since the
two vertex sets are disjoint. Let $W_h=W(G_h,\sigma_h)$ and put
$b=b_{h-1}$.

Any prefix of $X$ lies either within $A$ or consists of all of
$A$ and a prefix of $B$. Minimal transversals of a union of
hypergraphs on disjoint supports are exactly unions of one
minimal transversal from each side. Thus every prefix of $X$
has at most $bW_{h-1}$ minimal transversals, including the empty
prefix; the same holds for $Y$. Also $|\Tr(Y)|=b^2$.

Every nonfinal prefix of $X\star Y$ consists of all completed
rows indexed by a prefix $\mathcal R$ of $X$, followed by a
possibly empty prefix $\{e\}\star\mathcal S$ of the next row,
where $\mathcal S$ is a prefix of $Y$. Set
\[
 P=F_{\mathcal R},\qquad Q=F_Y,\qquad
 R=F_{\mathcal S},\qquad L_e=\bigvee_{u\in e}x_u.
\]
The CNF for the completed rows is $P\vee Q$, and that for the
partial row is $L_e\vee R$. These identities follow by
distributing a disjunction over the two CNFs; they remain valid
for empty prefixes, whose CNF is the constant one.
Since $\mathcal S\subseteq Y$, we have $Q\Rightarrow R$.
The full prefix function is therefore
\begin{equation}\label{berge:prefix-identity}
 (P\vee Q)\wedge(L_e\vee R)
       =Q\vee(P\wedge L_e)\vee(P\wedge R).
\end{equation}
If positive DNFs have $p$ and $q$ terms, their disjunction has
at most $p+q$ minimal terms and their conjunction has at most
$pq$: reduction can only decrease the number of generated terms.
Applying this to \eqref{berge:prefix-identity} gives the bound
\[
 b^2+(bW_{h-1})\bigl(2^{h-1}+bW_{h-1}\bigr)
\]
for the number of minimal transversals of the prefix.
The final prefix has $b_h=2b^2$ transversals and satisfies the
same eventual upper bound. Since
$2^{h-1}\le b=2^{2^{h-1}-1}$ and $W_{h-1}\ge1$, we conclude
\[
 W_h\le3b^2W_{h-1}^2.
\]
Writing $w_h=\log W_h$ and using $w_0=0$,
\[
 w_h\le2w_{h-1}+2(2^{h-1}-1)+\log3
       \le2w_{h-1}+2^h.
\]
Induction yields $w_h\le h2^h$.
Equation~\eqref{berge:takata-counts} gives
$\log N_h=\Theta(2^h)$ and $\log\log N_h=\Theta(h)$,
proving the width estimate. The time estimate follows from
the implementation bound preceding the proposition.
Combining the width upper bound with Takata's lower bound
proves the last assertion.
\end{proof}

In particular, no asymptotically stronger lower bound valid for
every edge order can be obtained on these same $G_h$. This does
not exclude larger intermediate families under unfavorable
orderings, or larger lower bounds on different hypergraphs.

\subsection{Single-edge deletion in the symmetric padding family}

Now use $\HH=\HH_{m,r}$ and $\LL=\Tr(\HH)$ from Section~4,
with $m\ge3$, $r\ge2$, and $m+1$ distinct evaluation points in
$\FF_q$. The family $\HH$ is reduced by Lemma~\ref{lem:Bcounts}.
For $E\in\HH$, define
\[
 \Delta_E=\Tr(\HH\setminus\{E\})\setminus\Tr(\HH).
\]
A minimal transversal $T$ of $\HH\setminus\{E\}$ belongs to
$\Delta_E$ if and only if $T\cap E=\varnothing$. Indeed, if it
also meets $E$, it is a transversal of $\HH$ and remains minimal:
any proper subset hitting $\HH$ would hit $\HH\setminus\{E\}$.
Thus $\Delta_E$ can be counted inside $U\setminus E$.

\begin{lemma}[Exact deletion counts]\label{berge:deletion}
Put $s=q^{r-1}$. For every padded edge,
\begin{equation}\label{berge:deletion-count}
 |\Delta_{E_{j,a}}|=
 \begin{cases}
 s^m,&j=0,\\[1mm]
 \displaystyle\frac{(m-1)(q-1)}{2q}s^m,&1\le j\le m.
 \end{cases}
\end{equation}
In particular, $|\Delta_E|\ge\tfrac12 q^{m(r-1)}$ for every
$E\in\HH$.
\end{lemma}
\begin{proof}
First let $E=E_{0,a}$. Its complement is the disjoint union of
\[
 D_\ell=\{(\ell,p):p(\pi_0)=a_\ell\},
 \qquad 1\le\ell\le m,
\]
each of size $s$. For a vector $b$ differing from $a$ only in
coordinate $\ell$, the edge $E_{0,b}$ is distinct from $E$ and
its intersection with $U\setminus E$ is exactly $D_\ell$.
Consequently any transversal of $\HH\setminus\{E\}$ contained
in $U\setminus E$ must meet every $D_\ell$.

Conversely, every set selecting one vertex from each $D_\ell$
meets all the remaining edges. An edge $E_{0,b}\ne E_{0,a}$
has a coordinate $\ell$ with $b_\ell\ne a_\ell$, and the
selected vertex of $D_\ell$ belongs to that edge. Every edge
of a block family $j\ge1$ contains $W_j$ and is met by the
selected vertex there. The edges $E_{0,b}$ differing in one
coordinate are private edges for these selected vertices.
The selectors are therefore minimal transversals. Any larger
set meeting every $D_\ell$ contains one of these selectors
and cannot be minimal. There are exactly $s^m$ such sets.

Next let $E=E_{i,a}$, where $i\ge1$. Its complement contains
no vertex in $W_i$ and is the union of
\[
 D_\ell=\{(\ell,p):p(\pi_i)=a_\ell\},
 \qquad \ell\ne i,
\]
again with $|D_\ell|=s$. For $\ell\ne i$, changing only
$a_\ell$ in the indexing vector produces another family-$i$
edge whose intersection with $U\setminus E$ is $D_\ell$.
Thus a transversal in this complement must meet every remaining
block. This condition also suffices to hit every remaining
family-$i$ edge and every block-family edge with index other
than $i$.

It remains to hit the empty-base family $0$. By
Lemma~\ref{lem:onefamily}, a set in the complement hits that
family exactly when, in some $D_\ell$, it contains two vertices
with different evaluations at $\pi_0$. Any transversal in the
complement therefore contains a set consisting of such a pair
in one block and a singleton in every other remaining block.
Every set of this form hits all edges except $E$.
It is minimal: removing a singleton empties a required block;
removing either member of the pair leaves at most one vertex
in each block, which cannot hit the empty-base family.
Hence these are exactly the minimal transversals missing $E$.

There are $m-1$ choices for the doubled block. Because
$\pi_0\ne\pi_i$ and $r\ge2$, prescribing the two evaluations
of a degree-less-than-$r$ polynomial is a surjective linear
map with fibers of size $q^{r-2}=s/q$. Within a fixed
$D_\ell$, the number of unordered pairs with different
$\pi_0$-evaluations is consequently
\[
 \binom q2(s/q)^2=\frac{q-1}{2q}s^2.
\]
The other $m-2$ blocks contribute $s^{m-2}$ singleton choices.
Multiplying gives the second case of
\eqref{berge:deletion-count}. Distinct choices give distinct
sets, since the doubled block and the pair can be recovered
from the set. Finally $q\ge m+1\ge4$ implies
$(m-1)(q-1)/(2q)\ge1/2$, proving the uniform bound.
\end{proof}

\begin{theorem}[An order-independent lower bound for Berge multiplication]
\label{berge:theorem}
Let $k=4t^2$ and let $q$ be the smallest power of two at least
$k+1$, as in Theorem~\ref{thm:B}. Every edge ordering of
$\HH_{k,t}$ forces ordinary Berge multiplication to generate
a penultimate intermediate family with at least
\[
 \frac12 q^{k(t-1)}
   =N^{\Omega(\sqrt{\log N/\log\log N})}
\]
minimal transversals, where $N$ is the combined input--output
incidence size. In particular, its running time is at least
this large. Explicit storage of the whole intermediate family
requires at least this many set records.
\end{theorem}
\begin{proof}
Whatever edge is processed last, say $E$, the preceding prefix
is exactly $\HH_{k,t}\setminus\{E\}$. By
Lemma~\ref{berge:deletion}, its transversal family has at least
$|\Delta_E|\ge\tfrac12 q^{k(t-1)}$ members. This applies to
every ordering without assumptions on the ordering heuristic.

We recall the size estimates to make the conversion explicit.
With $v=kq^t$, Lemmas~\ref{lem:Bcounts} and \ref{lem:Bsmall} give
\[
 |\HH_{k,t}|=q^k+kq^{k-1},\qquad
 |\Tr(\HH_{k,t})|\le(v+1)^{2t+1},\qquad
 |T|\le2t\quad(T\in\Tr(\HH_{k,t})).
\]
Therefore
\[
 q^k\le N\le2+v+(v+1)(q^k+kq^{k-1})
                       +(2t+1)(v+1)^{2t+1}.
\]
Since $k=4t^2$ and $q=\Theta(t^2)$, these inequalities yield
$\log N=\Theta(t^2\log t)$ and
$\log\log N=\Theta(\log t)$. On the other hand,
\[
 \log\bigl(\tfrac12q^{k(t-1)}\bigr)
       =k(t-1)\log q-1=\Omega(t^3\log t).
\]
Thus the logarithm of the intermediate-family lower bound is
$\Omega(\log N\sqrt{\log N/\log\log N})$, as claimed.
The same conclusion holds for the standard binary encoding,
whose logarithm differs by at most $O(\log\log(v+2))$.
\end{proof}

The exponent $\sqrt{\log N/\log\log N}$ grows faster than
$\log\log N$, so Theorem~\ref{berge:theorem} strengthens the
order-independent bound supplied by Takata's family. It uses
the symmetric specialization of the KKW padding already
analyzed for Algorithm B, but a different obstruction:
every possible last edge exposes a large family of new minimal
transversals immediately before it is processed. No ETH
assumption and no decision-tree lower bound is used.

\section{Scope, explicitness, and concluding remarks}

\paragraph{Summary of the results.}
We proved two lower bounds on the number of recursive calls of the original
Fredman--Khachiyan procedures, as a function of the explicit input size $N$.
For Algorithm A, every execution on the read-once pairs $(\HH_k,\LL_k)$ of Gurvich and
Khachiyan makes $N^{\Omega(\log N)}$ calls (Theorem~\ref{gka:theorem}); the argument
bounds the size of every decision tree computing $f_k$ through the evaluation of its
multilinear polynomial at $z=(1+i\sqrt3)/2$, so it holds for every branching-variable choice
and for the whole class $\mathcal R_\alpha$ of two-child variable-restriction procedures with
$\alpha(N)=2^{o((\log N)^2)}$. On this family the count is in fact $N^{\Theta(\log N)}$.
For Algorithm B, every execution on the symmetric padded pairs $(\HH_{k,t},\LL_{k,t})$ makes
$N^{\Omega(\sqrt{\log N/\log\log N})}$ calls (Theorem~\ref{thm:B}).
The same padded family forces every ordering of Berge multiplication
to generate at least $\tfrac12q^{k(t-1)}$ minimal transversals in
its penultimate intermediate family, giving the same exponent
scale in the combined input--output size
(Theorem~\ref{berge:theorem}). For Takata's read-once family,
Proposition~\ref{berge:takata-upper} instead shows that the
$N^{\Omega(\log\log N)}$ bound is tight in exponent scale
when the ordering is chosen optimally.
Applying the padding construction of \cite{KKW26} directly to Algorithm A, as done in the
appendix (Theorem~\ref{thm:A}), only gives a bound in the same exponent scale as that for B;
the read-once family is what yields the stronger bound for A.
The known upper bounds are $N^{O(\log^2 N)}$ for Algorithm A and $N^{O(\log N/\log\log N)}$
for Algorithm B, so gaps remain in both cases. Since the number of calls of any admissible
execution of A on the family $(\HH_k,\LL_k)$ is already $N^{\Theta(\log N)}$, an improvement of the
lower bound for A beyond $N^{O(\log N)}$ would require different input families.
Closing the remaining gaps is left open.

\paragraph{Explicitness.}
All input families are finite and explicit, and no unavailable transversal oracle is used.
The pairs $(\HH_k,\LL_k)$ are generated directly from the recursion
\eqref{gka:recursion} and its dual by unions and Cartesian products on disjoint supports, in time
polynomial in their explicit size $N_k$.
For the B family, the polynomial labels and padded edges can be enumerated directly, and the
transversal family can be computed by enumerating subsets up to the proved size bound and testing
hitting and minimality. This takes $2^{O(t^2\log t)}$ time, including redundancy removal: the number
of candidates, the number of edges, and the incidence-processing cost each have a logarithm
$O(t^2\log t)$.
In the B family the empty original edge is padded before reduction; reducing the
un-padded family first would define a different construction.
Because $q$ can be a power of two, a full polynomial block can also be identified with the
assignments to $t\log q$ Boolean variables.

\paragraph{Scope.}
The FK-A and FK-B proofs are separate and use different families. The Algorithm B theorem is stated
directly for its explicit hypergraphs; it does not assert the same result for every SAT formula
or for every truncated labeling in the construction of the appendix.
Likewise, the read-once bound does not transfer to Algorithm B, whose secondary
decompositions are not the two-child restriction recursion required by $\mathcal R_\alpha$; Remark~\ref{rem:B-decision-tree} gives an explicit example.
The bounds for the original algorithms hold for every permitted variable choice in
the recursion-tree model of Remark~\ref{def:model}. The generalized variable-restriction
bounds allow the terminal rules specified in Definition~\ref{def:restriction-class}.
Appendix~\ref{rosep:section} additionally allows read-once terminal functions and proves
that the weaker KKW-based bound survives, although the Gurvich--Khachiyan family
can then be accepted at the root. None of these bounds covers sharing isomorphic
subproblems or arbitrary alternative decompositions without further proof.
The Berge bound concerns explicit generation of complete prefix transversal families,
and does not apply to algorithms that avoid these intermediate families.
The arguments require neither ETH nor SETH, and no hardness of the source unsatisfiable
formula is invoked.

\paragraph{AI disclosure statement.}
The author acknowledges the use of generative AI technologies in the preparation of this paper. Specifically, ChatGPT (OpenAI) was utilized iteratively to push the initial ideas into complete proofs, while Gemini (Google) and Claude (Anthropic) were used to refine those proofs, clarify arguments, and improve the overall writing quality. The author takes full responsibility for the correctness and integrity of the final results.

\bibliographystyle{alpha}
\bibliography{dualn}

\appendix

\section{A weaker lower bound for variable-restriction algorithms based on KKW}\label{sec:weakLB}

\subsection{A selector-tree lower bound}
We first isolate a lower-bound principle that is independent of padding. At its core, Lemma~\ref{lem:equality} establishes a tight combinatorial lower bound on the number of final outcomes (leaves) that a binary decision tree must possess if it attempts to shrink large coordinate spaces down to a target size using only point-equality queries. It considers a decision tree that operates on a product space of independent coordinates, starting at the root with the domain \(D^{\text{root}} = D_1 \times \cdots \times D_k\), where \(\vert{}D_i\vert{} = d_i\). At each internal node, the tree executes an equality query of the form:
\[ \text{Does coordinate } i \text{ equal the specific value } c? \]
This split divides the current product domain D into two parallel branches: in the ``Yes'' branch, the choice for coordinate $i$ instantly collapses to a singleton, $|D^{\text{yes}}_i| = 1$. This coordinate is successfully solved, and all other coordinate domains remain unchanged.
In the ``No'' branch, the choice for coordinate $i$ only shrinks by a single element, leaving $|D^{\text{no}}_i| = |D_i| - 1$. All other coordinate domains remain unchanged.

\begin{lemma}[Equality-query tree]\label{lem:equality}
Let $d_i\ge s_i\ge1$ be integers. Consider a binary decision tree on independent
coordinates with initial domains of sizes $d_1,\ldots,d_k$, whose queries
ask whether one coordinate equals a specified value. Suppose each
nonempty terminal domain is a product $D_1\times\cdots\times D_k$
with $|D_i|\le s_i$. The tree has at least
\[
 \prod_{i=1}^k(d_i-s_i+1)
\]
nonempty leaves.
\end{lemma}
\begin{proof}
We track the structural complexity of the product domains as they are partitioned down the decision tree by defining a potential weight function. For any nonempty product domain \(D = D_1 \times \cdots \times D_k\), we assign the weight:
\[w(D)=w(D_1,\ldots,D_k) = \prod_{m=1}^k \max\{1,\vert{}D_m\vert{}-s_m+1\}.\]
Suppose an internal node queries the \(i\)-th coordinate, asking whether its value equals a specified constant \(c\). This query splits the active parent domain \(D\) into two potential child domains, \(D^{\text{yes}}\) and \(D^{\text{no}}\). Since the coordinates are independent, the domains for all other coordinates \(m \neq i\) remain completely unchanged in both children. Let \(K\) denote the constant product of these unaffected factors:
\[K = \prod_{m \neq i} \max\{1,\vert{}D_m\vert{}-s_m+1\}.\]
Because each component factor is a positive integer, we have \(K \ge 1\). Let \(a = \vert{}D_i\vert{}\) be the size of the queried coordinate domain prior to the split. If \(a \ge 2\) and \(c\in D_i\), the query divides \(D_i\) into a ``Yes'' branch of size $1$ (containing only $\{c\}$) and a ``No'' branch of size $a-1$ (containing $D_i \setminus \{c\}$). The sum of the weights of the two children can be calculated by factoring out $K$:
\[
 w(D^{\text{yes}}) + w(D^{\text{no}}) = K \cdot \Big[ \max\{1, 2-s_i\} + \max\{1, a-s_i\} \Big].
\]
We analyze this sum based on how the parent domain size $a$ relates to the threshold $s_i$:
\begin{itemize}
    \item \textbf{Case 1 ($a > s_i$):} Since $s_i \ge 1$, we have $2-s_i \le 1$, meaning $\max\{1, 2-s_i\} = 1$. Because $a - s_i \ge 1$, we have $\max\{1, a-s_i\} = a-s_i$. Substituting these yields:
    \[
     w(D^{\text{yes}}) + w(D^{\text{no}}) = K \cdot [1 + (a-s_i)] = K \cdot (a-s_i+1) = w(D).
    \]
    Thus, the total weight is perfectly conserved under the split.
    \item \textbf{Case 2 ($2 \le a \le s_i$):} Here, we have $2-s_i \le 1$ and $a-s_i \le 0$, which means both terms inside the brackets collapse to $1$. This yields:
    \[
     w(D^{\text{yes}}) + w(D^{\text{no}}) = K \cdot [1 + 1] = K \cdot 2 \ge K \cdot 1 = w(D),
     \]
     since the parent factor was $\max\{1, a-s_i+1\} = 1$. Thus, the total weight strictly increases.
\end{itemize}
If $a=1$, the sole feasible outcome leaves the domain and its weight unchanged. If the queried value $c$ does not belong to the current domain $D_i$, there is only one feasible outcome (the ``No'' branch), leaving the domain and its weight unchanged. In all scenarios, the sum of the nonempty child weights is at least the parent weight.

Repeated application of this local inequality down the finite tree guarantees that the initial root weight is bounded above by the total sum of the leaf weights:
\[w(D^{\text{root}}) \le \sum_{L \in \text{Leaves}} w(D^L).\]
At the root, the initial domain sizes are \(\vert{}D_i\vert{} = d_i \ge s_i\), meaning \(w(D^{\text{root}}) = \prod_{i=1}^k (d_i-s_i+1)\). At any nonempty terminal leaf \(L\), the stopping criterion ensures that \(\vert{}D_i\vert{} \le s_i\) for all \(i \in [k]\), causing every coordinate factor to collapse to \(1\). Thus, each nonempty leaf has a terminal weight of \(w(D^L) = 1\). Substituting these values into the summation yields:
\[\prod_{i=1}^k (d_i-s_i+1) \le \sum_{L \in \text{Leaves}} 1 = \vert{}\text{Nonempty Leaves}\vert{},\]
which completes the proof.
\end{proof}

\begin{proposition}[General selector lower bound]\label{prop:A}
Let $U=W_1\dot\cup\cdots\dot\cup W_k$, with $|W_i|=d_i$.
Suppose $\HH$ is the hypergraph defined in Section~\ref{sec:polypad}, $\LL=\Tr(\HH)$, no full selector is a
transversal, and for integers $1\le s_i\le d_i$,
\begin{equation}\label{eq:smallcomplement}
 |W_i\setminus E|\le s_i
 \quad\hbox{for every }E\in\HH\hbox{ and every }i.
\end{equation}
Every execution of a procedure in $\mathcal R_\alpha$ on $(\HH,\LL)$
of incidence size $N$ has at least
\[
 \frac{1}{\alpha(N)}\prod_{i=1}^k(d_i-s_i+1)
\]
terminal invocations, and hence at least that many calls.
For original Algorithm A one may take $\alpha(N)=2$.
\end{proposition}

\begin{proof}
Let $L$ be the number of terminal invocations.
Lemma~\ref{lem:restriction-completion} completes the execution to a
Boolean decision tree computing $F_{\HH}$ with at most $\alpha(N)L$
leaves.

Restrict its inputs to characteristic vectors of full selectors. By the proposition's hypothesis, no full selector is a valid transversal of $\HH$. Since $F_{\HH}(\mathbf{1}_X) = 1$ if and only if \(X\) contains a transversal, it follows that every full selector $X\subseteq U$ satisfies $F_{\HH}(\mathbf{1}_X)=0$. At any reachable node, the
possible selected vertices form a product $D_1\times\cdots\times D_k$, where initially $D_i=W_i$.
Querying $x_v$ for $v\in W_i$ asks whether the selected vertex in block
$i$ equals $v$; the two possible new domains are $\{v\}$ and
$D_i\setminus\{v\}$, when nonempty. Since the blocks are disjoint, fixing or shrinking the domain of block \(i\) has no impact on the choices available in any other block. Thus, the restricted tree functions exactly as an equality-query decision tree operating on independent coordinates with initial domain sizes \(\vert{}W_i\vert{} = d_i\).

At a zero leaf of the full Boolean decision tree, let $Z$ be the set
of vertices whose variables were fixed to zero. Some original edge
$E\in\HH$ must satisfy $E\subseteq Z$. Otherwise, setting all remaining
unassigned variables outside of \(Z\) to one would meet every original edge and give a
true input extending that leaf. Every possible selected vertex at this
leaf avoids $Z$, so
\[
 D_i\subseteq W_i\setminus Z\subseteq W_i\setminus E.
\]
By \eqref{eq:smallcomplement}, $|D_i|\le s_i$.
Lemma~\ref{lem:equality} now shows that the expanded decision tree has at
least $\prod_i(d_i-s_i+1)$ selector-containing leaves. Its total number
of leaves is at most $\alpha(N)L$, proving the bound.
No frequency condition or other restriction on variable selection was used.
\end{proof}

\subsection{A balanced labeling}

Recall that an evaluation fiber at a fixed point $\pi \in \FF_q$ for a target value $v \in \FF_q$ and a set of polynomials $p_a\in\FF_q[x]$, $a \in \{0, \ldots, d-1\}$, is the set of all integers whose associated polynomials evaluate to $v$:
\[
\text{Fiber}(\pi, v) = \{a \in \{0, \ldots, d-1\} : p_a(\pi) = v\}.
\]
\begin{lemma}[Simultaneously balanced evaluation fibers]\label{lem:balanced}
Suppose $q$ is prime and $d\le q^t$. The first $d$ nonnegative integers
give an injective labeling by degree-less-than-$t$ polynomials such that
every evaluation fiber at every point of $\FF_q$ has size at most
$\lceil d/q\rceil$.
\end{lemma}
\begin{proof}
Write each integer $0\le a<d$ in base $q$ as
$a=c_0+c_1q+\cdots+c_{t-1}q^{t-1}$ and associate
$p_a(z)=c_0+c_1z+\cdots+c_{t-1}z^{t-1}$.
Different integers have different coefficient vectors, ensuring that the polynomial assignment $a \mapsto p_a(z)$ is strictly injective.
Consider a complete group of $q$ consecutive integers beginning at a multiple
of $q$:
\[
G_B = \{B \cdot q, \, B \cdot q + 1, \, \ldots, \, B \cdot q + (q-1)\},
\]
where $B \ge 0$ is a fixed integer block index.
For every integer $a \in G_B$, the higher-order base-$q$ digits $(c_1, c_2, \ldots, c_{t-1})$ are entirely determined by the quotient $B = \lfloor a/q \rfloor$, meaning they remain perfectly constant across all elements within $G_B$. Meanwhile, the least significant digit $c_0 = a \pmod q$ runs over the entire field $\FF_q$ exactly once.

Evaluating the polynomials of $G_B$ at an arbitrary fixed point $\pi \in \FF_q$ yields:
\[
p_a(\pi) = c_0 + \sum_{m=1}^{t-1} c_m \pi^m = c_0 + C_B(\pi),
\]
where $C_B(\pi) = \sum_{m=1}^{t-1} c_m \pi^m$ is a fixed field element constant for all of $G_B$. Because the mapping $c_0 \mapsto c_0 + C_B(\pi)$ is a translation and therefore a bijection over $\FF_q$, the total evaluation $p_a(\pi)$ sweeps across every element of the field exactly once as $c_0$ ranges from $0$ to $q-1$. Consequently, each complete group $G_B$ contributes exactly one element to any chosen fiber $\text{Fiber}(\pi, v)$.

The domain $\{0, 1, \ldots, d-1\}$ contains exactly $\lfloor d/q \rfloor$ complete groups of size $q$. If $d$ is not a multiple of $q$, there remains a final, incomplete group containing the last $d \pmod q$ elements. 

Within this final incomplete segment, the higher coefficients are still fixed, and the constant coefficients $c_0$ are strictly distinct. Since $c_0$ does not cover the full field, the shifted evaluations $c_0 + C_B(\pi)$ are distinct but incomplete, meaning any target value $v \in \FF_q$ can appear at most once.

Summing the contributions from all segments, the total number of times any value $v$ can appear as an evaluation outcome at a point $\pi$ is bounded by:
\[
|\text{Fiber}(\pi, v)| \le \lfloor d/q \rfloor + \mathbf{1}_{q\nmid d} = \lceil d/q \rceil.
\]
\end{proof}

With the labeling of Lemma~\ref{lem:balanced} in each block, \eqref{eq:padding} immediately gives
\[
 W_i\setminus E_{j,a}\subseteq\{u\in W_i:h_j(u)=a_i\},
 \qquad
 |W_i\setminus E_{j,a}|\le\left\lceil d_i/q\right\rceil.
\]
These inequalities survive reduction (of computing $\HH = \Min(\HH_{\rm raw})$) because every retained edge is an
original padded edge.

\subsection{Application to the SAT construction}
We recall the construction in~\cite{KKW26}.
Let $\varphi$ be an unsatisfiable formula of clause width at most three
on a specified set of $n$ variables. Duplicate clauses can be removed,
leaving $O(n^3)$ clauses. Partition the variables into $k$ nonempty
blocks $B_i$ of size at most $b$, and let $W_i= \left\{ (i, \beta) : \beta \in \{0,1\}^{|B_i|} \right\}$ consist of their
$2^{|B_i|}$ assignments. In addition to the original block edges $W_i$,
include for each clause $c$ the edge
\[
 E'_c=\{(i,\beta):\text{some literal of }c\text{ in }B_i
                         \text{ is true under block assignment }\beta\}.
\]
A full selector meets all these original edges exactly when its assembled
assignment satisfies $\varphi$: a clause edge is met exactly when that
clause has a true literal. Since $\varphi$ is unsatisfiable, there is no
transversal full selector.

For $n\ge32$ use the parameter choices
\[
 \ell=\lceil\log(n+1)\rceil,\quad
 t=\max\{z\ge1:4z^3\ell\le n\},\quad
 b=t\ell,\quad k=\lceil n/b\rceil .
\]
Choose a balanced partition with block size at most $b$.
Let $M$ be the number of original edges, put $Q=\max\{M,2^\ell\}$,
and choose a prime $q$ with $Q\le q<2Q$.
Then $q=n^{O(1)}$, $q\ge n+1$, and
\[
 |W_i|\le2^b\le q^t,\qquad
 k\ge n/(t\ell)\ge4t^2>2t.
\]
Label the assignment vertices by Lemma~\ref{lem:balanced} and form the
padded hypergraph. By Lemma~\ref{lem:small}, all its minimal transversals
have size at most $2t$. Thus the family of small minimal transversals given to the algorithm is the entire family $\LL=\Tr(\HH)$.

\begin{theorem}[KKW-based lower bound for variable-restriction algorithms]
\label{thm:A}
On the preceding pair obtained from an unsatisfiable $n$-variable
formula, every execution of a procedure in $\mathcal R_\alpha$ has at
least
\[
 \frac{2^n}{e\,\alpha(N)}
\]
terminal invocations, for arbitrary variable selection.
In particular, original Algorithm A has at least $2^n/(2e)$ such
invocations. For every fixed procedure with polynomial terminal
decision-tree complexity, its worst-case call count along the explicit
infinite family below is at least
\[
 N^{\Omega(\sqrt{\log N/\log\log N})}.
\]
\end{theorem}
\begin{proof}
Apply Proposition~\ref{prop:A} with
$d_i=2^{|B_i|}$ and $s_i=\lceil d_i/q\rceil$. Since
$d_i-\lceil d_i/q\rceil+1\ge d_i(1-1/q)$ and
$\prod_i d_i=2^n$, the number of leaves is at least
\[
 \frac{2^n}{\alpha(N)}\,(1-1/q)^k.
\]
Because $q\ge n+1$ and $k\le n$,
\[
 (1-1/q)^k\ge(n/(n+1))^n\ge e^{-1}.
\]
For the last inequality use $(1+1/n)^n\le e$.

It remains to express the bound in terms of the explicit input.
The definitions give
\[
 t=\Theta((n/\log n)^{1/3}),\quad
 b=\Theta(n^{1/3}(\log n)^{2/3}),\quad
 k=\Theta((n/\log n)^{2/3}).
\]
Writing $v=|U|$, the edge and transversal counts satisfy
\[
 v\le k2^b,\qquad
 |\HH|\le Mq^k,\qquad
 |\LL|\le (v+1)^{2t+1}.
\]
Multiplying the first edge count by at most $v$ incidences per edge,
and the second by at most $2t$ incidences per transversal, shows
\[
 \log N=O\bigl(n^{2/3}(\log n)^{1/3}\bigr).
\]
The same holds for a binary encoding. Set $s=n^{2/3}(\log n)^{1/3}$.
The function $x^{3/2}/\sqrt{\log x}$ is increasing for sufficiently large
$x$, so
\[
 \frac{(\log N)^{3/2}}{\sqrt{\log\log N}}
 =O\left(\frac{s^{3/2}}{\sqrt{\log s}}\right)=O(n).
\]
For a fixed procedure with $\alpha(N)\le2N^C$, the leaf lower bound
has logarithm at least $n-C\log N-O(1)$. The displayed upper bound
on $\log N$ is $o(n)$, so this logarithm is $\Omega(n)$.
Together with the preceding comparison this proves the assertion.
More generally, the same conclusion holds whenever
$\log\alpha(N)=o(n)$ along this family.
An explicit infinite source family is
$\varphi_n=(x_1)\land(\neg x_1)\land\bigwedge_{i=2}^n(x_i)$.
Every variable occurs, and these formulas have width at most three.
\end{proof}

\section{A separation obtained by allowing read-once terminal functions}
\label{rosep:section}

The bounded-terminal-complexity model can be enlarged in a way that
distinguishes the two constructions. Allow a procedure to terminate
when the residual common Boolean function is monotone read-once,
even if that function has a large decision tree. The selector lower
bound from the KKW construction survives this additional stopping
rule. In contrast, the Gurvich--Khachiyan functions themselves are
read-once, so a procedure that always uses this stopping rule accepts
those dual inputs at the root.

We regard constants as read-once. A monotone read-once function is one
represented by an AND--OR formula in which every variable appears at
most once. Restricting variables to constants preserves this property:
substitute the constants and simplify the resulting formula.

\begin{definition}[Variable restriction with read-once stopping]
\label{rosep:class}
For $\alpha(N)\ge1$, let $\mathcal R_{\mathrm{RO},\alpha}$ consist of
correct duality-testing procedures satisfying conditions (i), (ii),
and (iv) of Definition~\ref{def:restriction-class}, with condition
(iii) replaced by the following requirement. On a dual input of
incidence size $N$, the common Boolean function at every terminal
invocation is either
\begin{enumerate}[label=(\alph*)]
\item monotone read-once; or
\item computable by a variable-query decision tree with at most
$\alpha(N)$ leaves.
\end{enumerate}
The size $N$ is that of the original input. Both restrictions are
executed at every nonterminal invocation, and invocations are counted
separately, without memoization. A procedure may use any sound method
to verify its terminal cases. Define
$\mathcal R^{\mathrm{stop}}_{\mathrm{RO},\alpha}$ to be the subclass
that always terminates whenever the residual dual pair represents a
read-once function.
\end{definition}

These are semantic requirements on executions on dual inputs; they
do not permit a procedure to assume duality without checking it.
We give a polynomial-time implementation of the read-once terminal
test below. First we prove that allowing these terminals does not
destroy the selector argument.

\subsection{A zero certificate for read-once functions}

For a variable set $D$ and $u\in D$, let $e_u$ denote the assignment
that is one at $u$ and zero at every other variable of $D$.
A zero certificate is a set $C\subseteq D$ such that assigning zero
to every variable in $C$ forces the function to be zero, regardless
of the values of the other variables.

\begin{lemma}[A small zero certificate]\label{rosep:certificate}
Let $h$ be a monotone read-once function on a nonempty set $D$ of
$d$ variables, some of which may be inessential. If
$h(e_u)=0$ for every $u\in D$, then $h$ has a zero certificate
of size at most $d/2$.
\end{lemma}
\begin{proof}
The constant-zero case has the empty certificate. The constant-one
case is excluded by the hypothesis. For a nonconstant function,
remove inessential variables and simplify a read-once formula so
that its leaves are variables and every internal gate has at least
two children. It suffices to prove the bound with the number of
essential variables, since this is at most $d$.

A formula consisting of one variable violates the hypothesis.
If the root is an AND gate, choose a child using the fewest variables.
Its variable set has size at most half the total. Setting all its
variables to zero makes that nonconstant monotone child zero and
therefore makes the whole conjunction zero.

If the root is an OR gate, each child is zero on every singleton
assignment to its own variables; otherwise the whole formula would
be one on such a singleton. Inductively, each child has a zero
certificate using at most half its variables. The children have
disjoint variable sets. Taking the union of their certificates
forces all children to zero and uses at most half the total variables.
This completes the induction.
\end{proof}

\begin{lemma}[Selector domains at a read-once node]
\label{rosep:domains}
Suppose $U=W_1\dot\cup\cdots\dot\cup W_k$, no full selector is a
transversal of $\HH$, and
\[
 |W_i\setminus E|\le s_i
 \qquad(E\in\HH,\ 1\le i\le k),
 \qquad s_i\ge1.
\]
Let $\rho$ be a partial assignment compatible with at least one full
selector. Write $D_1\times\cdots\times D_k$ for the nonempty product
of possible selected vertices compatible with $\rho$.
If $F_{\HH}|_\rho$ is read-once, then
\[
 |D_i|\le2s_i\qquad(1\le i\le k).
\]
\end{lemma}
\begin{proof}
Fix $i$, and write $d=|D_i|$. The conclusion is immediate if $d=1$.
Suppose $d\ge2$. No variable in block $W_i$ can have been assigned
one by $\rho$, since that would fix the selected vertex and leave
only one possibility. The variables of $D_i$ are all unassigned.

For each $j\ne i$, choose any $w_j\in D_j$. Fix the variables of
$W_j$ to the singleton assignment selecting $w_j$, and fix all
variables of $W_i\setminus D_i$ to zero. These assignments extend
$\rho$ on every variable outside $D_i$. The resulting function $h$
on $D_i$ is a restriction of the read-once function $F_{\HH}|_\rho$,
and hence is read-once. For every $u\in D_i$, the assignment $e_u$
completes these fixed choices to a full selector. Thus $h(e_u)=0$.

By Lemma~\ref{rosep:certificate}, there is a zero certificate
$C\subseteq D_i$ with $|C|\le d/2$. Assign zero to $C$ and one to
all of $D_i\setminus C$, retaining the assignments outside $D_i$.
The resulting full Boolean assignment makes $F_{\HH}$ zero.
Consequently some original edge $E\in\HH$ is missed by its true
support. Since every vertex of $D_i\setminus C$ was set to one,
$E\cap D_i\subseteq C$. On the other hand, the original complement
bound implies
\[
 d-s_i\le |E\cap D_i|\le |C|\le d/2.
\]
Therefore $d\le2s_i$. This argument applies independently to each
block, proving the assertion.
\end{proof}

\subsection{The extended selector lower bound}

\begin{proposition}[Selector lower bound with read-once stopping]
\label{rosep:general}
Let $U=W_1\dot\cup\cdots\dot\cup W_k$, with $|W_i|=d_i$.
Assume $\HH$ is reduced, $\LL=\Tr(\HH)$, no full selector is a
transversal, and $|W_i\setminus E|\le s_i$ for all $E\in\HH$,
where $1\le s_i\le d_i$. Every execution in
$\mathcal R_{\mathrm{RO},\alpha}$ on this pair has at least
\begin{equation}\label{rosep:product}
 \frac1{\alpha(N)}
 \prod_{i=1}^k\max\{1,d_i-2s_i+1\}
\end{equation}
terminal invocations, where $N$ is the incidence size in
\eqref{eq:size}.
\end{proposition}
\begin{proof}
Let $L$ be the number of terminal invocations. At a read-once terminal,
make no change. At any other terminal, attach a decision tree of
at most $\alpha(N)$ leaves computing the residual function. Its leaves
represent constant residual functions, which are read-once.
This produces a variable-query tree with at most $\alpha(N)L$ leaves,
all of whose residual functions are read-once. It is not necessary
to evaluate the read-once functions at these leaves.

Restrict the inputs of this tree to full selectors. Each query to
$x_v$ for $v\in W_i$ asks whether the chosen vertex of block $i$
equals $v$. The feasible choices at any reachable node form a
nonempty product domain. At every reachable leaf,
Lemma~\ref{rosep:domains} bounds its $i$th domain size by
$\tau_i:=\min\{d_i,2s_i\}$.
Apply Lemma~\ref{lem:equality} with these $\tau_i$. The number of
selector-containing leaves is at least
\[
 \prod_i(d_i-\tau_i+1)=\prod_i\max\{1,d_i-2s_i+1\}.
\]
It is at most $\alpha(N)L$, proving \eqref{rosep:product}.
\end{proof}

\begin{theorem}[KKW bound despite read-once stopping]
\label{rosep:kkw}
For the balanced KKW-based construction in Theorem~\ref{thm:A},
every execution of a procedure in $\mathcal R_{\mathrm{RO},\alpha}$
has at least
\begin{equation}\label{rosep:kkw-exact}
 \frac1{\alpha(N)}\prod_{i=1}^k
 \max\left\{1,d_i-2\left\lceil\frac{d_i}{q}\right\rceil+1\right\}
 \ \ge\ \frac{2^{n-2k}}{\alpha(N)}
\end{equation}
terminal invocations, for all sufficiently large $n$.
For every fixed procedure with $\alpha(N)=N^{O(1)}$, its call count
along the explicit infinite family of Theorem~\ref{thm:A} is at least
\[
 N^{\Omega(\sqrt{\log N/\log\log N})}.
\]
This includes the subclass
$\mathcal R^{\mathrm{stop}}_{\mathrm{RO},\alpha}$ that always
terminates on read-once residual dual pairs.
\end{theorem}
\begin{proof}
In that construction, $d_i=2^{|B_i|}\ge2$, $\prod_i d_i=2^n$,
and the balanced fibers give
$|W_i\setminus E|\le\lceil d_i/q\rceil$ for every original reduced
padded edge. Proposition~\ref{rosep:general} gives the product in
\eqref{rosep:kkw-exact}. For sufficiently large $n$, $q\ge n+1\ge8$.
Using $\lceil d/q\rceil\le d/q+1$, we obtain, for every $d\ge2$,
\[
 d-2\lceil d/q\rceil+1
 \ge d(1-2/q)-1
 \ge\frac{3d}{4}-1
 \ge\frac d4.
\]
Multiplication gives the second inequality in
\eqref{rosep:kkw-exact}.

The parameter estimates already proved for Theorem~\ref{thm:A} are
\[
 k=\Theta((n/\log n)^{2/3})=o(n),\qquad
 \log N=O(n^{2/3}(\log n)^{1/3})=o(n).
\]
Thus, for $\alpha(N)=N^{O(1)}$, the logarithm of the leaf lower bound
is $n-2k-\log\alpha(N)=\Omega(n)$. The same size conversion as in
Theorem~\ref{thm:A} gives the asserted bound in terms of $N$.
\end{proof}

\subsection{Why the read-once hard family no longer gives a lower bound}

\begin{proposition}[Separation of the two families]
\label{rosep:separation}
There is a correct variable-restriction duality algorithm in
$\mathcal R^{\mathrm{stop}}_{\mathrm{RO},1}$ whose read-once
terminal test runs in polynomial time in the current explicit input.
On every dual pair representing a Gurvich--Khachiyan read-once function,
it makes exactly one recursive invocation and runs in polynomial time.
Nevertheless, on the balanced KKW family it satisfies the
superpolynomial lower bound of Theorem~\ref{rosep:kkw}.
\end{proposition}
\begin{proof}
At each invocation, remove redundant terms and clauses. Recognize
whether the monotone function given by the DNF family $\LL$ is
read-once and, if so, obtain a read-once formula for it.
This can be done in polynomial time from its prime DNF
\cite{GMR06}. For the read-once case, use Gurvich's
read-once dualization method, based on the read-once structure
(see the treatment by Golumbic and Gurvich in
\cite[Chapter~10]{GG11}).

If the DNF function is read-once, test whether its canonical CNF is
exactly the supplied reduced family $\HH$. This comparison is also
polynomial in the current input size. To see this, first simplify
constants in the read-once formula. For a nonconstant formula, the
number of prime clauses is the sum of the child counts at an AND
gate and their product at an OR gate. Compute these counts with
saturation at $|\HH|+1$. If the root count differs from $|\HH|$,
the pair is not dual. Otherwise every subformula has at most
$|\HH|$ prime clauses: all child counts are positive and sums and
products dominate each factor. Generate the clauses bottom-up and
compare them with $\HH$. Disjoint variable sets ensure that the
generated clauses are distinct and minimal. There are at most
polynomially many clauses to process across the formula, each of
length at most the current number of variables. Constants are checked
directly. The test therefore accepts exactly the dual pairs in this
read-once case.

If the DNF function is not read-once, choose any remaining variable
and recursively test both restrictions, accepting precisely when
both are dual. The restriction equivalence proves correctness; the
number of remaining variables decreases, so the procedure terminates.
On dual inputs, its only terminal functions are read-once and it
always stops on them. Hence it belongs to
$\mathcal R^{\mathrm{stop}}_{\mathrm{RO},1}$.

The Gurvich--Khachiyan functions
\[
 f_1=x_1\vee x_2,\qquad
 f_{j+1}=(f_j^{(1)}\wedge f_j^{(2)})
          \vee(f_j^{(3)}\wedge f_j^{(4)})
\]
use disjoint copies and are read-once by construction. The root test
accepts their dual pairs, so there is one invocation and polynomial
work. The KKW assertion follows from Theorem~\ref{rosep:kkw} with
$\alpha=1$.
\end{proof}

\begin{remark}[The distinction between the two lower-bound arguments]
The earlier completion argument bounds the decision-tree size of every
terminal function. It fails when a complicated read-once function may
be accepted at a terminal: read-once formula size does not bound
decision-tree size polynomially. The proof above instead bounds the
selector domains at each read-once terminal directly. Its key change
is $s_i\mapsto2s_i$, which costs only $2^{O(k)}$ in the KKW parameters.
Thus this is a separation for a natural structural stopping rule,
not a test designed to recognize just one hard family.

The separation is between the specified constructions. It does not
rule out stronger lower bounds for this enlarged class on other
inputs. Nor does it include Algorithm B merely because its ordinary
terminal cases are simple: B's secondary decomposition does not
satisfy the required exhaustive two-child variable-restriction rule.
\end{remark}

\section{Full proofs of the lemmas from prior work}
\label{app:full-proofs}
For completeness, we give self-contained proofs of the restriction
lemmas from \cite{FK96} and the padding lemmas from \cite{KKW26}.
They are ordered by their appearance in the main text. Proofs of
the other main-text lemmas are given alongside their statements.

\subsection{Full proof of Lemma~\ref{lem:restrictions}}
\label{fullproof:lem:restrictions}
\begin{proof}
Write $U'=U\setminus\{v\}$. For $b\in\{0,1\}$ and
$y\in\{0,1\}^{U'}$, let $y^{(b)}$ be the extension with $x_v=b$.
Denote the raw clause and term families by
\begin{align*}
 \mathcal C_1&=\{E\in\HH:v\notin E\},&
 \mathcal D_1&=\{T\setminus\{v\}:T\in\LL\},\\
 \mathcal C_0&=\{E\setminus\{v\}:E\in\HH\},&
 \mathcal D_0&=\{T\in\LL:v\notin T\}.
\end{align*}
Substitution in the CNF gives
\[
 F(y^{(b)})=\bigwedge_{A\in\mathcal C_b}\bigvee_{u\in A}y_u.
\]
Indeed, setting $x_v=1$ satisfies precisely the clauses containing
$v$, while setting $x_v=0$ removes that literal from each clause.
Substitution in the DNF gives, for the same function,
\[
 F(y^{(b)})=\bigvee_{B\in\mathcal D_b}\bigwedge_{u\in B}y_u.
\]
Here setting $x_v=1$ removes the literal from every term containing
it, while setting $x_v=0$ makes every such term false.

Removing duplicates and supersets preserves each displayed function.
For clauses, a clause on a smaller vertex set implies one on a larger
set. For terms, a conjunction on a larger set implies the conjunction
on a smaller set, so the former is redundant in a disjunction.
Consequently the reduced representations are exactly
$\HH_{v=b}=\Min(\mathcal C_b)$ and
$\LL_{v=b}=\Min(\mathcal D_b)$ as stated.

For completeness, equality of these two reduced representations also
implies the required transversal identity. A true support for a
positive CNF is exactly a set meeting every clause. Thus its minimal
true supports are the minimal transversals of its clause family.
For a reduced positive DNF, its terms are exactly its minimal true
supports: every true support contains a term, and no term properly
contains another. Hence
$\LL_{v=b}=\Tr(\HH_{v=b})$ for $b=0,1$.
The empty-family and empty-set cases obey the same identities under
the conventions of Section~2, so no exceptional constant case is needed.
\end{proof}

\subsection{Full proof of Lemma~\ref{lem:onefamily}}
\label{fullproof:lem:onefamily}
\begin{proof}
\textbf{Sufficiency ($\Longleftarrow$):} 
Assume the condition holds. We show that $X \cap E_{j,a} \neq \varnothing$ for every target vector $a = (a_1,\ldots,a_k) \in \FF_q^k$.
If $X \cap E'_j \neq \varnothing$, then as $E'_j \subseteq E_{j,a}$ holds for all $a \in \FF_q^k$, $X \cap E_{j,a}\ne \varnothing$ for all $a$.
 Otherwise, there exists a block $W_i$ containing two vertices $u,v \in X$ such that $h_j(u) \neq h_j(v)$. For any choice of $a \in \FF_q^k$, consider its $i$-th component $a_i \in \FF_q$. Since $h_j(u)$ and $h_j(v)$ are distinct elements in $\FF_q$, the scalar $a_i$ can match at most one of them. Therefore, we must have $h_j(u) \neq a_i$ or $h_j(v) \neq a_i$, which guarantees that $u \in E_{j,a}$ or $v \in E_{j,a}$. Since $u,v \in X$, the set $X$ meets $E_{j,a}$ for all $a$.

\smallskip
\noindent
\textbf{Necessity ($\Longrightarrow$):} 
We prove the contrapositive. Suppose that $X \cap E'_j = \varnothing$ and that for every block $W_i$, all vertices in $X \cap W_i$ share a common evaluation value under $h_j$. We construct a specific vector $a = (a_1,\ldots,a_k) \in \FF_q^k$ as follows:
\begin{itemize}
    \item If $X \cap W_i \neq \varnothing$, let $c_i \in \FF_q$ be the unique value such that $h_j(u) = c_i$ for all $u \in X \cap W_i$. We set $a_i = c_i$.
    \item If $X \cap W_i = \varnothing$, we choose $a_i \in \FF_q$ arbitrarily.
\end{itemize}
We now verify the intersection $X \cap E_{j,a}$. Since $X \cap E'_j = \varnothing$, any element in the intersection must come from the padding. However, for any block $i \in [k]$ and any vertex $u \in X \cap W_i$, our construction ensures $h_j(u) = a_i$. Thus, no vertex in $X$ satisfies the condition $h_j(u) \neq a_i$. It follows that $X \cap E_{j,a} = \varnothing$, meaning $X$ fails to meet this specific edge.

\smallskip
\noindent
\textbf{Size Bound:} 
The case analysis in the sufficiency direction demonstrates that a small subset $Y \subseteq X$ with $|Y| \le 2$ can be explicitly chosen to hit the entire family. If $X \cap E'_j \neq \varnothing$, picking a single vertex $Y = \{u\}$ where $u \in X \cap E'_j$ suffices. Otherwise, picking the pair $Y = \{u,v\} \subseteq X \cap W_i$ with $h_j(u) \neq h_j(v)$ suffices. In either scenario, at most two vertices are required.
\end{proof}

\subsection{Full proof of Lemma~\ref{lem:small}}
\label{fullproof:lem:small}
\begin{proof}
\textbf{Selectors and Minimality:} 
First, observe that passing from $\HH_{\rm raw}$ to $\HH = \Min(\HH_{\rm raw})$ does not change the family of transversals, because every deleted edge contains a retained edge; thus, any set hitting a minimal edge automatically hits its supersets. 

Let $X \subseteq U$ be a set containing at most one vertex per block (i.e., $|X \cap W_i| \le 1$ for all $i \in [k]$). Lemma~\ref{lem:onefamily} implies that meeting all padded families $\mathcal{F}_j = \{E_{j,a} : a \in \FF_q^k\}$ is equivalent to meeting every base edge $E'_j \in \HH'$. Since the original family of base edges $\HH'$ explicitly included the blocks $W_1,\ldots,W_k$, the set $X$ must select exactly one vertex from each block if $X$ is a transversal of $\HH$. Hence, $X$ is a full selector of size exactly $k$. Removing any chosen vertex from $X$ causes it to miss that block's original edge. Lemma~\ref{lem:onefamily} then shows that it misses at least one edge of the corresponding padded family, proving minimality.

\smallskip
\noindent
\textbf{Transversals with Multi-Element Blocks:} 
Now, let $X$ be a minimal transversal of $\HH$, and suppose there exists some block $W_i$ containing two distinct vertices $u,v \in X$. By construction, these vertices are labeled by distinct polynomials $p_u, p_v \in \FF_q[x]$ of degree strictly less than $t$. The difference polynomial $P(x) = p_u(x) - p_v(x)$ is a non-zero polynomial of degree at most $t-1$. Let $I$ denote the index set of padded families where $u$ and $v$ yield identical evaluations:
\[
 I = \{j \in [M] : h_j(u) = h_j(v)\} = \{j \in [M] : p_u(\pi_j) = p_v(\pi_j)\}.
\]
Because the evaluation points $\pi_1, \ldots, \pi_M$ are pairwise distinct, each $j \in I$ corresponds to a distinct root of $P(x)$ in $\FF_q$. By the polynomial root bound over a field, a non-zero polynomial of degree at most $t-1$ has at most $t-1$ roots, which yields the bound $|I| \le t-1$.

We now analyze the hitting properties of the pair $\{u,v\} \subseteq X$:
\begin{itemize}
    \item For any family indexed outside $I$ (where $j \notin I$), we have $h_j(u) \neq h_j(v)$. The proof of Lemma~\ref{lem:onefamily} guarantees that the pair $\{u,v\}$ automatically meets every edge in the family $\mathcal{F}_j$.
    \item For each exceptional family $j \in I$, Lemma~\ref{lem:onefamily} implies that we can select a small subset $Y_j \subseteq X$ containing at most two vertices that suffices to meet every edge in $\mathcal{F}_j$.
\end{itemize}
Consider the combined hitting set $Y = \{u,v\} \cup \bigcup_{j \in I} Y_j$. By construction, $Y \subseteq X$ forms a valid transversal for the entire hypergraph $\HH$. We can bound the size of this set explicitly:
\[
 |Y| \le 2 + \sum_{j \in I} |Y_j| \le 2 + 2|I| \le 2 + 2(t-1) = 2t.
\]
Since $X$ is assumed to be a \emph{minimal} transversal, it cannot contain any proper subset that is also a transversal. Because $Y \subseteq X$ is a valid transversal, it must hold that $X = Y$, establishing that $|X| \le 2t$.
\end{proof}

\subsection{Full proof of Lemma~\ref{lem:shift}}
\label{fullproof:lem:shift}
\begin{proof}
Let $x$ be an arbitrary vertex in $\tau_g(T \cap E)$. By the definition of an image set, there exists a vertex $v \in T \cap E$ such that $\tau_g(v) = x$. Since $v \in T \cap E$, it follows that $x \in \tau_g(T) \cap \tau_g(E)$. Thus, $\tau_g(T \cap E) \subseteq \tau_g(T) \cap \tau_g(E)$.
For the reverse direction, let $x$ be an arbitrary vertex in $\tau_g(T) \cap \tau_g(E)$. Then there exist vertices $v_1 \in T$ and $v_2 \in E$ such that $\tau_g(v_1) = \tau_g(v_2) = x$.
We now invoke the injectivity of the induced mapping. Let $v_1 = (i_1, p_1)$ and $v_2 = (i_2, p_2)$. The condition $\tau_g(v_1) = \tau_g(v_2)$ expands to:
\(
(i_1, p_1 + g) = (i_2, p_2 + g).
\)
Equating coordinates forces $i_1 = i_2$ and $p_1 + g = p_2 + g$. Canceling $g$ shows $p_1 = p_2$, meaning $v_1 = v_2$. Let this unique input vertex be denoted by $v$. Because $v \in T$ and $v \in E$, we have $v \in T \cap E$. Therefore, its image $x = \tau_g(v)$ must belong to the image of the intersection, giving $x \in \tau_g(T \cap E)$. Thus, $\tau_g(T) \cap \tau_g(E) \subseteq \tau_g(T \cap E)$.
Combining both directions yields the desired set identity $\tau_g(T) \cap \tau_g(E) = \tau_g(T \cap E)$.

Now we prove that $\tau_g(\HH_{m,r})$ is a permutation of $\HH_{m,r}$.
Recall from \eqref{eq:padding2} that a padded edge $E_{j,a} \in \HH_{m,r}$ is defined by a index $j$ and a vector $a \in \FF_q^m$. Applying the map elementwise gives $\tau_g(E_{j,a}) = \{(i, p+g) : (i,p) \in E_{j,a}\}$. The base edge $E'_j$ is either empty or a whole block and is preserved by $\tau_g$. Outside $E'_j$, a vertex $(i,p+g)$ lies in the image if and only if $p(\pi_j)\ne a_i$, which can be rearranged as:
\[
(p+g)(\pi_j) - g(\pi_j) \neq a_i \iff (p+g)(\pi_j) \neq a_i + g(\pi_j).
\]
Defining a transformed target vector $a' \in \FF_q^m$ where $a'_i = a_i + g(\pi_j)$ for all $i$, this condition yields exactly the padded edge $E_{j,a'}$. Thus, $\tau_g(E_{j,a}) = E_{j,a'}$, meaning $\tau_g$ maps every edge in $\HH_{m,r}$ to another valid edge within the same block family $j$. Because the shifting mapping $a \mapsto a'$ is a bijection over $\FF_q^m$, $\tau_g$ maps $\HH_{m,r}$ injectively into itself. Since $\HH_{m,r}$ is a finite set, this injective mapping is a full permutation of $\HH_{m,r}$.

Finally, we prove that $\tau_g(\LL_{m,r})$ is a permutation of $\LL_{m,r}$.
Let $T \in \LL_{m,r}$ be a minimal transversal. By definition, $T \cap E \neq \varnothing$ for all $E \in \HH_{m,r}$. Applying our proven intersection identity, we have:
\[
\tau_g(T) \cap \tau_g(E) = \tau_g(T \cap E) \neq \varnothing.
\]
Because $\tau_g$ permutes $\HH_{m,r}$, as $E$ ranges over all edges of $\HH_{m,r}$, its image $\tau_g(E)$ also ranges over all edges of $\HH_{m,r}$. Thus, $\tau_g(T)$ intersects every edge in the hypergraph, proving it is a valid transversal.

To establish minimality, note that since $T$ is minimal, every vertex $v \in T$ has a critical edge $E_v \in \HH_{m,r}$ such that $T \cap E_v = \{v\}$. Applying the intersection identity yields:
\[
\tau_g(T) \cap \tau_g(E_v) = \tau_g(T \cap E_v) = \{\tau_g(v)\}.
\]
Since $\tau_g(E_v) \in \HH_{m,r}$, the vertex $\tau_g(v)$ is unique in tracking that edge. Removing $\tau_g(v)$ would leave $\tau_g(E_v)$ completely un-intersected. Thus, no vertex can be removed from $\tau_g(T)$ without violating the transversal property, making $\tau_g(T)$ a minimal transversal ($\tau_g(T) \in \LL_{m,r}$). Finally, since $\tau_g$ maps the finite set $\LL_{m,r}$ injectively into itself via an invertible operation (with inverse $\tau_{-g}$), it follows that $\tau_g(\LL_{m,r})$ is a permutation of $\LL_{m,r}$.
\end{proof}

\subsection{Full proof of Lemma~\ref{lem:secondary}}
\label{fullproof:lem:secondary}
\begin{proof}
We first identify the secondary verification directly from the two
DNF representations used by FK96. Let $y$ denote the variables other
than $x_v$, and define
\begin{align*}
 a(y)&=\bigvee_{\substack{T\in\LL\\v\in T}}
                   \bigwedge_{u\in T\setminus\{v\}}y_u,&
 b(y)&=\bigvee_{\substack{T\in\LL\\v\notin T}}
                   \bigwedge_{u\in T}y_u,\\
 c(y)&=\bigvee_{\substack{A\in\HH\\v\in A}}
                   \bigwedge_{u\in A\setminus\{v\}}y_u,&
 d(y)&=\bigvee_{\substack{A\in\HH\\v\notin A}}
                   \bigwedge_{u\in A}y_u.
\end{align*}
Then the common function is $x_va(y)\vee b(y)$, and its monotone
dual is $x_vc(y)\vee d(y)$. The main call for $x_v=0$ establishes
\begin{equation}\label{eq:secondary-main-identity}
 b(y)=1-\bigl(c(\bar y)\vee d(\bar y)\bigr).
\end{equation}
The remaining duality identity is
$a(y)\vee b(y)=1-d(\bar y)$.
Because every $T\in\LL$ meets every $A\in\HH$, we have
$a(y)\wedge d(\bar y)=0$: a term defining $a$ and an edge defining
$d$ share some variable other than $v$, and cannot both be true at
$y$ and $\bar y$, respectively.

If $c(\bar y)=0$, equation~\eqref{eq:secondary-main-identity} gives
$b(y)=1-d(\bar y)$, so the remaining identity already holds.
If $c(\bar y)=1$, it gives $b(y)=0$, and the remaining identity
reduces to $a(y)=1-d(\bar y)$. Now $c(\bar y)=1$ is the union of
the conditions $y_u=0$ for all $u\in J$, where
$J=E\setminus\{v\}$ and $v\in E\in\HH$.
Thus the secondary calls of the rule in Section~4 of \cite{FK96}
test the duality of $a|_{J=0}$ and $d|_{J=1}$, one call for each
such $E$. In the DNF $d$, substituting one on $J$ removes $J$
from every term. In $a$, substituting zero on $J$ deletes every
term meeting $J$. Their reduced term families are precisely
\eqref{eq:secH} and \eqref{eq:secL}, respectively, in that order.
Notice that the opposite substitutions occur because the arguments
of the two dual DNFs are complementary.

Finally, we verify the claimed interpretation as a restriction of
the original common Boolean function. Set $x_v=1$ and set $x_u=0$
for $u\in J$. In the CNF, all clauses containing $v$ disappear;
each other clause $A$ becomes $A\setminus J$. Its reduced clause
family is therefore \eqref{eq:secH}. In the DNF, a term $T$ meeting
$J$ vanishes. If $v\notin T$, the transversal property and
$E=J\cup\{v\}$ imply $T\cap J\ne\varnothing$, so every such term
vanishes as well. The surviving terms contain $v$, avoid $J$,
and lose the literal $x_v$, giving exactly \eqref{eq:secL} after
reduction. The two families represent the same restricted function;
by the minimal-true-support argument in the proof of
Lemma~\ref{lem:restrictions}, the latter family is the transversal
family of the former. This proves both the formula for the call
and its functional interpretation, including constant residuals.
\end{proof}

\end{document}